\documentclass{aa} %[referee]
\usepackage[varg]{txfonts}
\usepackage{soul}
\usepackage{ifthen}
\usepackage{graphicx}
\usepackage{bbm}
\usepackage{Paper_Linc}
\usepackage[AA]{Math_Linc}
\defcitealias{Fan_etal_2010}{FSL10}
\defcitealias{Liu_etal_2015}{LPH15}
\defcitealias{Liu_etal_2015a}{LPL15}
\defcitealias{Lin_Kilbinger_2015}{Paper I}
\newcommand{\abd}{\mathrm{abd}}
\newcommand{\pct}{\mathrm{pct}}
\newcommand{\cut}{\mathrm{cut}}
\newcommand{\true}{\mathrm{true}}

\title{A new model to predict weak-lensing peak counts}
\subtitle{II. Parameter constraint strategies}
\titlerunning{A new model to predict weak-lensing peak counts II.}
\author{Chieh-An Lin \and Martin Kilbinger}
\authorrunning{C.-A. Lin \& M. Kilbinger}
\institute{
	Service d'Astrophysique, CEA Saclay, Orme des Merisiers, B\^at 709, 91191 Gif-sur-Yvette, France\\
	\email{\texttt{chieh-an.lin@cea.fr}}
} 
\date{Received 2 June 2015 / Accepted 17 August 2015}

\abstract%
	{Peak counts have been shown to be an excellent tool for extracting the non-Gaussian part of the weak lensing signal. Recently, we developed a fast stochastic forward model to predict weak-lensing peak counts. Our model is able to reconstruct the underlying distribution of observables for analysis.}
	{In this work, we explore and compare various strategies for constraining a parameter using our model, focusing on the matter density $\Omega_\mathrm{m}$ and the density fluctuation amplitude $\sigma_8$.}
	{First, we examine the impact from the cosmological dependency of covariances (CDC). Second, we perform the analysis with the copula likelihood, a technique that makes a weaker assumption than does the Gaussian likelihood. Third, direct, non-analytic parameter estimations are applied using the full information of the distribution. Fourth, we obtain constraints with approximate Bayesian computation (ABC), an efficient, robust, and likelihood-free algorithm based on accept-reject sampling.}
	{We find that neglecting the CDC effect enlarges parameter contours by 22\% and that the covariance-varying copula likelihood is a very good approximation to the true likelihood. The direct techniques work well in spite of noisier contours. Concerning ABC, the iterative process converges quickly to a posterior distribution that is in excellent agreement with results from our other analyses. The time cost for ABC is reduced by two orders of magnitude.}
	{The stochastic nature of our weak-lensing peak count model allows us to use various techniques that approach the true underlying probability distribution of observables, without making simplifying assumptions. Our work can be generalized to other observables where forward simulations provide samples of the underlying distribution.}
	

\begin{document}
\maketitle

\section{Introduction}
\label{sect:intro}

Weak lensing (WL) is a gravitational deflection effect of light by
matter inhomogeneities in the Universe that causes distortion of source
galaxy images. This distortion corresponds to the integrated deflection
along the line of sight, and {its measurement} probes the high-mass regions of
the Universe. These regions contain structures that formed during the
late-time evolution of the Universe, which depends on cosmological
parameters, such as the matter density parameter $\Omegam$, the matter density
fluctuation $\sigeig$, and the equation of state of dark energy
$w$. Ongoing and future surveys such as KiDS \footnote{\url{http://kids.strw.leidenuniv.nl/}},
DES \footnote{\url{http://www.darkenergysurvey.org/}},
HSC \footnote{\url{http://www.naoj.org/Projects/HSC/HSCProject.html}},
WFIRST \footnote{\url{http://wfirst.gsfc.nasa.gov/}}, 
Euclid \footnote{\url{http://www.euclid-ec.org/}}, 
and LSST \footnote{\url{http://www.lsst.org/lsst/}} 
are expected to provide tight constraints on those and other cosmological parameters and
to distinguish between different cosmological models, using weak lensing as a major probe.

Lensing signals can be extracted in several ways. A common observable
is the cosmic shear two-point-correlation function (2PCF), which has been used to 
constrain cosmological parameters in many studies, including recent ones
\citep{Kilbinger_etal_2013, Jee_etal_2013}.
However, the
2PCF only retains Gaussianity, and it misses the rich nonlinear information of
the structure evolution encoded on small scales. To compensate for this drawback,
several non-Gaussian statistics have been proposed, for example higher order moments
\citep{Kilbinger_Schneider_2005, Semboloni_etal_2011, Fu_etal_2014, Simon_etal_2015}, 
the three-point correlation function \citep{Schneider_Lombardi_2003, Takada_Jain_2003, Scoccimarro_etal_2004},
Minkowski functionals \citep{Petri_etal_2015},
or peak statistics, which is the aim of this series of papers.
Some more general work comparing different 
strategies to extract non-Gaussian information can be found in the
literature \citep{Pires_etal_2009a, Berge_etal_2010, Pires_etal_2012}.

Peaks, defined as local maxima of the lensing signal, are direct tracers
of high-mass regions in the large-scale structure of the Universe. 
In the medium and high signal-to-noise (S/N) regime, the peak function (the number of peaks as function of S/N)
is not dominated by shape noise, 
and this permits one to study the cosmological dependency of the
peak number counts \citep{Jain_VanWaerbeke_2000}.
Various aspects of peak statistic have
been investiagated in the past: 
the physical origin of peaks \mbox{\citep{Hamana_etal_2004, Yang_etal_2011}}, 
projection effects \citep{Marian_etal_2010}, 
the optimal combination of angular scales \citep{Kratochvil_etal_2010, Marian_etal_2012}, 
redshift tomography \citep{Hennawi_Spergel_2005},
cosmological parameter constraints \citep{Dietrich_Hartlap_2010, Liu_etal_2014}, 
detecting primordial non-Gaussianity \citep{Maturi_etal_2011, Marian_etal_2011}, 
peak statistics beyond the abundance \citep{Marian_etal_2013},
the impact from baryons \citep{Yang_etal_2013, Osato_etal_2015},
magnification bias \citep{Liu_etal_2014a}, 
and shape measurement errors \citep{Bard_etal_2013}.
Recent studies by 
\defcitealias{Liu_etal_2015}{Liu, Petri, Haiman et al.}\citetalias{Liu_etal_2015} 
(\citeyear{Liu_etal_2015}, hereafter\defcitealias{Liu_etal_2015}{LPH15}\citetalias{Liu_etal_2015}),
\defcitealias{Liu_etal_2015a}{Liu, Pan, Li et al.}\citetalias{Liu_etal_2015a} 
(\citeyear{Liu_etal_2015a}, hereafter\defcitealias{Liu_etal_2015a}{LPL15}\citetalias{Liu_etal_2015a}),
and \citet{Hamana_etal_2015}
have applied likelihood estimation for WL peaks on real data 
and have shown that the results agree with the current $\Lambda$CDM scenario.

Modeling number counts is a challenge for peak studies. 
To date, there have been three main approaches. 
The first one is to count peaks from a large
number of $N$-body simulations (\citealt{Dietrich_Hartlap_2010};
\citetalias{Liu_etal_2015}), 
which directly emulate structure formation by numerical implementation of the corresponding
physical laws.
The second family consists of analytic predictions 
\citep{Maturi_etal_2010, Fan_etal_2010} based on 
Gaussian random field theory. 
A third approach has been introduced by \citet[][hereafter
\citetalias{Lin_Kilbinger_2015}]{Lin_Kilbinger_2015}:
Similar to \citet{Kruse_Schneider_1999} and 
\citet{Kainulainen_Marra_2009, Kainulainen_Marra_2011, Kainulainen_Marra_2011a}, 
we propose a stochastic process to predict peak counts by simulating 
lensing maps from a halo distribution drawn from the mass function.

Our model possesses several advantages. 
The first one is flexibility. 
Observational conditions can easily be modeled and taken into account. 
The same is true for additional features, such as 
intrinsic alignment of galaxies and other observational 
and astrophysical systematics. 
Second, since our method does not need $N$-body simulations, 
the computation time required to calculate the model are orders of
magnitudes faster, and we can explore a large parameter space.
Third, our model explores the underlying 
probability density function (PDF) of the observables. All
statistical properties of the peak function can be derived directly from the model,
making various parameter estimation methods possible.

In this paper, we apply several parameter constraint and likelihood
methods for our peak-count-prediction model from
\citetalias{Lin_Kilbinger_2015}. Our goal is to
study and compare different strategies and to make use of
the full potential of the fast stochastic forward modeling approach. We
start with a likelihood function that is assumed to be Gaussian in the
observables with constant covariance and then compare this to methods that make fewer
and fewer assumptions, as follows.

The first extension of the Gaussian likelihood is to take
the cosmology-dependent covariances \citep[CDC, see][]{Eifler_etal_2009} into account. 
Thanks to the fast performance of our model, it is feasible to estimate the 
covariance matrix for each parameter set.

The second improvement we adopt is the copula analysis 
\citep{Benabed_etal_2009, Jiang_etal_2009, Takeuchi_2010, Scherrer_etal_2010, Sato_etal_2011}
for the Gaussian approximation.
Widely used in finance, the copula transform uses the fact that any
multivariate distribution can be transformed into a new one where the marginal
PDF is uniform. Combining successive transforms can then give rise to a
new distribution where all marginals are Gaussian. This makes weaker
assumptions about the underlying likelihood than the Gaussian hypothesis.

Third, we directly estimate the full underlying distribution
information in a non-analytical way. This allows us to strictly follow the
original definition of the likelihood estimator: the conditional probability of
observables for a given parameter set. In addition, we compute
the $p$-value from the full PDF. These $p$-values derived for
all parameter sets allow for significance tests and provide a direct way
to construct confidence contours.

Furthermore, our model makes it possible to dispose of a likelihood
function altogether, using approximate Bayesian computation (ABC, see e.g.
\citealt{Marin_etal_2011}) for exploring the parameter space. ABC is a
powerful constraining technique based on accept-reject sampling. Proposed first
by \citet{Rubin_1984}, ABC produces the posterior distribution by bypassing the
likelihood evaluation, which may be complex and practically unfeasible in some
contexts. The posterior is constructed by comparing the sampled result with the
observation to decide whether a proposed parameter is accepted. This technique
can be improved by combining ABC with population Monte Carlo (PMC 
\footnote{This algorithm is called PMC ABC by some and SMC (sequential Monte Carlo) ABC by others.},
\citealt{Beaumont_etal_2009, Cameron_Pettitt_2012, Weyant_etal_2013}). 
Until now, ABC seems to already have various applications in
biology-related domains \citep[e.g.,][]{Beaumont_etal_2009, Berger_etal_2010,
Csillery_etal_2010, Drovandi_Pettitt_2011}, while applications for astronomical
purposes are few: morphological transformation of galaxies 
\citep{Cameron_Pettitt_2012}, cosmological parameter inference using type Ia
supernovae \citep{Weyant_etal_2013}, constraints of the disk
formation of the Milky Way \citep{Robin_etal_2014}, 
and strong lensing properties of galaxy clusters \citep{Killedar_etal_2015}. Very
recently, two papers \citep{Ishida_etal_2015, Akeret_etal_2015} dedicated to
ABC in a general cosmological context have been submitted.

The paper is organized as follows. In \sect{sect:methodology}, we
briefly review our model introduced in \citetalias{Lin_Kilbinger_2015},
the setting for the parameter analysis, and the criteria for defining
parameter constraints. In
\sect{sect:CDC}, we study the impact of the CDC effect.
The results from the copula likelihood can be found in \sect{sect:copula}, and
in \sect{sect:nonParam} we estimate the true underlying PDF in a
non-analytic way and show parameters constraints without the Gaussian
hypothesis. \sect{sect:ABC} focuses on the likelihood-free ABC technique,
and the last section is dedicated to a discussion where we summarize this
work.

\section{Methodology}
\label{sect:methodology}

\subsection{Our model}
\label{sect:methodology_model}

Our peak-count model uses a probabilistic approach that generates peak catalogs from a given mass function model. This is done by generating fast simulations of halos, computing the projected mass, and simulating lensing maps from which one can extract WL peaks. A step-by-step summary is given as follows:
\begin{enumerate}
        \item sample halo masses and assign density profiles and positions (fast simulations),
        \item compute the projected mass and subtract the mean over the field (ray-tracing simulations),
        \item add noise and smooth the map with a kernel, and
        \item select local S/N maxima.
\end{enumerate}

Here, two assumptions have been made: (1) only bound matter contributes to number counts and (2) the spatial correlation of halos has a small impact on WL peaks. \citetalias{Lin_Kilbinger_2015} showed that combining both hypotheses gives a good estimation of the peak abundance.

\begin{table}
	\centering
	\caption{List of parameter values adopted in this study.}
	\begin{tabular}{ccc}
		\hline\hline\\[-2ex]
		Parameter                        & Symbol            & Value\\
		\hline\\[-1.8ex]
		Lower sampling limit             & -                 & $\dix{12}$ $\Msol/h$\\
		Upper sampling limit             & -                 & $\dix{17}$ $\Msol/h$\\
		NFW inner slope                  & $\alpha$          & 1\\
		$M$-$c$ relation parameter       & $c_0$             & 11\\
		$M$-$c$ relation parameter       & $\beta$           & 0.13\\
		Source redshift                  & $z_\rms$          & 1\\
		Intrinsic ellipticity dispersion & $\sigma_\epsilon$ & 0.4\\
		Galaxy number density            & $n_\rmg$          & 25 arcmin$\invSq$\\
		Pixel size                       & $\theta_\pix$     & 0.2 arcmin\\
		Kernel size                      & $\theta_\rmG$     & 1 arcmin\\
		Shape noise                      & $\sigma_\pix$     & 0.283\\
		Smoothed noise                   & $\sigma_\noise$   & 0.0226\\
		Effective field area             & -                 & 25 deg$^2$\\
		\hline
	\end{tabular}
	\label{tab:parameters}
\end{table}

We adopt the same settings as \citetalias{Lin_Kilbinger_2015}: the mass function model from \cite{Jenkins_etal_2001}, the truncated Navarro-Frenk-White halo profiles \citep{Navarro_etal_1996, Navarro_etal_1997}, Gaussian shape noise, the Gaussian smoothing kernel, and sources at fixed redshift which are distributed on a regular grid. The field of view is chosen such that the effective area after cutting off the border is 25 deg$^2$. An exhausted list of parameter values used in this paper can be found in \tab{tab:parameters}. Readers are encouraged to read \citetalias{Lin_Kilbinger_2015} for their definitions and for detailed explanations for our model.

All computations with our model in this study are performed by our \textsc{Camelus} algorithm~\footnote{\url{http://github.com/Linc-tw/camelus}}. A realization (from a mass function to a peak catalog) of a 25-deg$^2$ field costs few seconds to generate on a single-CPU computer. The real time cost depends of course on input cosmological parameters, but this still gives an idea about the speed of our algorithm.

\subsection{Analysis design}
\label{sect:methodology_design}

\begin{figure}[tb]
	\centering
	\includegraphics[width=7cm]{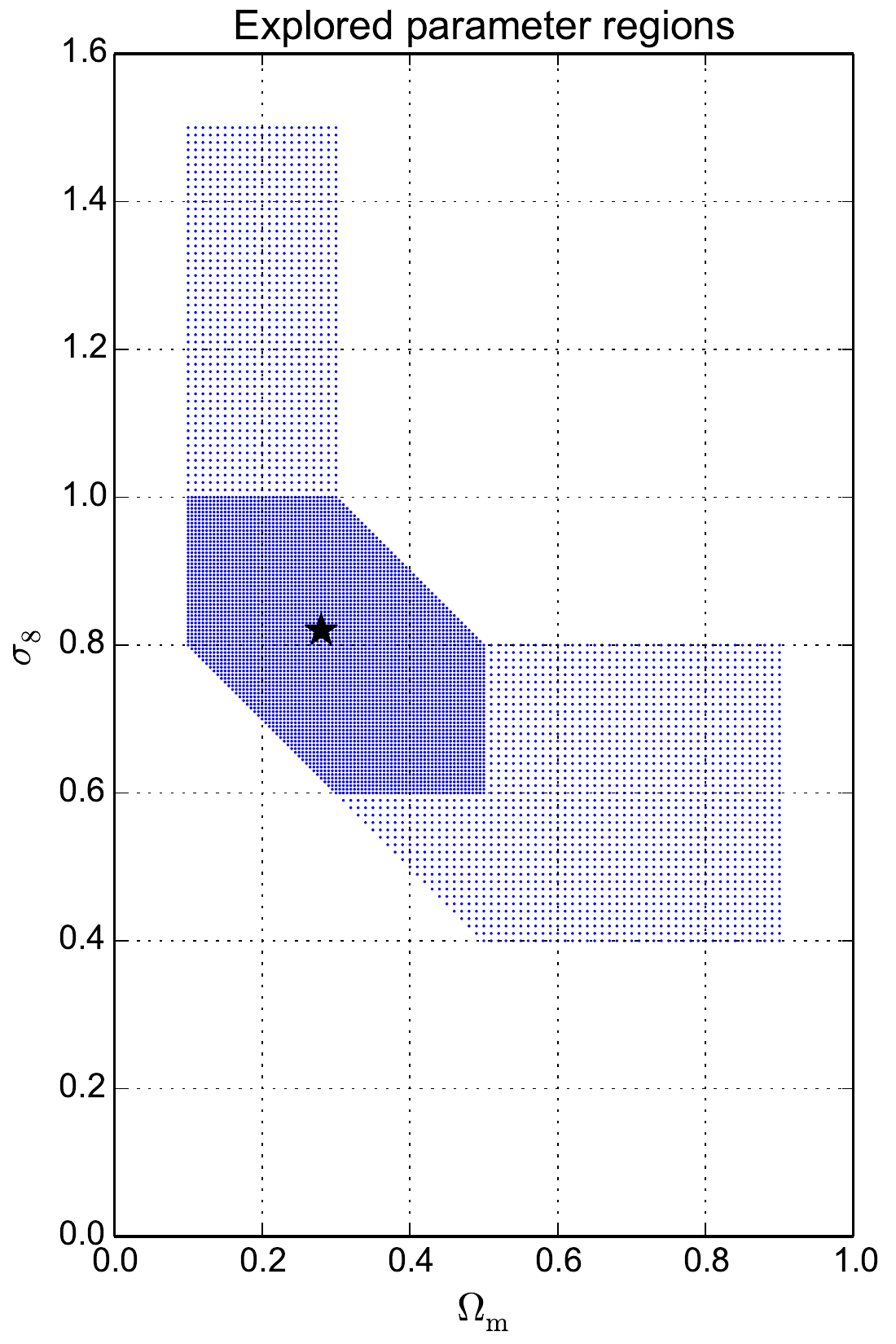}
        \caption{Location of 7821 points on which we evaluate the likelihoods. In the condensed area, the interval between two grid points is 0.005, while in both wing zones it is 0.01. The black star shows $(\Omegam^\inp, \sigeig^\inp) = (0.28, 0.82)$.}
	\label{fig:explored_regions}
\end{figure}

Throughout this paper, $\bpi$ denotes a parameter set. To simplify the study, the dimension of the parameter space is reduced to two: $\bpi \equiv (\Omegam, \sigeig)$. The other cosmological parameters are fixed, including $\Omega_\rmb = 0.047$, $h = 0.78$, $n_s = 0.95$, and $w=-1$. The dark energy density $\Omega_{\rmd\rme}$ is set to $1-\Omegam$ to match a flat universe. On the $\Omegam$-$\sigeig$ plane, we explore a region where the posterior density, or probability, is high, see \fig{fig:explored_regions}. We compute the values of three different log likelihoods on the grid points of these zones. The grid size of the center zone is $\Delta\Omegam = \Delta\sigeig = 0.005$, whereas it is 0.01 for the rest. This results in a total of 7821 points in the parameter space to evaluate.

For each $\bpi$, we carry out $N = 1000$ realizations of a 25-deg$^2$ field and determine the associated data vector $\bx\upp{k} = (x\upp{k}_1, \ldots, x\upp{k}_d)$ for all $k$ from 1 to $N$. These are independent samples drawn from their underlying PDF of observables for a given parameter $\bpi$. We estimate the model prediction (which is the mean), the covariance matrix, and the inverse matrix \citep{Hartlap_etal_2007}, respectively, by following
\begin{align}
	x^\model_i &= \frac{1}{N}\sum_{k=1}^N x_i^{(k)}, \label{for:mean_estimator}\\
	\hat{C}_{ij} &= \frac{1}{N-1}\sum_{k=1}^N \left(x_i^{(k)} - x^\model_i\right) \left(x_j^{(k)} - x^\model_j\right), \ \ \text{and}  \label{for:cov_estimator}\\
	\widehat{\bC\inv} &= \frac{N-d-2}{N-1}\ \widehat{\bC}\inv, \label{for:invCov_estimator}
\end{align}
where $d$ denotes the dimension of data vector. This results in a total area of 25~000 deg$^2$ for the mean estimation.

In this paper, the observation data $\bx^\obs$ are identified with a realization of our model, which means that $\bx^\obs$ is derived by a particular realization of $\bx(\bpi^\inp)$. The input parameters chosen are $\bpi^\inp = (\Omegam^\inp, \sigeig^\inp) = (0.28, 0.82)$. The authors would like to highlight that the accuracy of the model is not the aim of this research work, but precision. Therefore, the input choice and the uncertainty of random process should have little impact.

Peak-count information can be combined into a data vector using different ways. Inspired by \cite{Dietrich_Hartlap_2010} and \citetalias{Liu_etal_2015a}, we studied three types of observables. The first is the abundance of peaks found in each S/N bin (labeled \texttt{abd}), in other words, the binned peak function. The second is the S/N values at some given percentiles of the peak cumulative distribution function (CDF, labeled \texttt{pct}). The third is similar to the second type, but without taking peaks below a threshold S/N value (labeled \texttt{cut}) into account. Mathematically, the two last types of observables can be denoted as $x_i$, thereby satisfying
\begin{align}\label{for:definition_percentile}
	p_i = \int_{\nu_\minn}^{x_i} n_\peak(\nu)\rmd\nu,
\end{align}
where $n_\peak(\nu)$ is the peak PDF function, $\nu_\minn$ a cutoff, and $p_i$ a given percentile. The observable $\bx^\abd$ is used by \citetalias{Liu_etal_2015a}, while readers find $\bx^\cut$ from \cite{Dietrich_Hartlap_2010}. We would like to clarify that using $\bx^\pct$ for analysis could by risky, since this includes peaks with negative S/N. From \citetalias{Lin_Kilbinger_2015}, we observe that although high-peak counts from our model agree well with $N$-body simulations, predictions for local maxima found in underdensity regions (peaks with S/N <~0) are inaccurate. Thus, we include $\bx^\pct$ in this paper only to give an idea about how much information we can extract from observables defined by percentiles.

\begin{table}
	\centering
	\caption{Definition of $\bx^{\abd 5}$, $\bx^{\pct 5}$, and $\bx^{\cut 5}$. Parameters such as $\nu_\minn$ and $p_i$ are used by \for{for:definition_percentile}. As an indication, their values for the input cosmology $\bpi^\inp$ are also given. They were calculated by averaging over 1000 realizations.}
	\label{for:pct_like_observables}
	\begin{tabular}{cc@{\hspace*{0.6em}}c@{\hspace*{0.6em}}c@{\hspace*{0.6em}}c@{\hspace*{0.6em}}c}
		\hline\hline\\[-1.8ex]
		Label                & \texttt{abd5}\\
		Bins on $\nu$        & \tiny[3.0, 3.8[ & \tiny[3.8, 4.5[ & \tiny[4.5, 5.3[ & \tiny[5.3, 6.2[ & \tiny[6.2, $+\infty$[\\
		$x_i$ for $\pi^\inp$ & 330   & 91    & 39    & 18    & 15\\
		\hline\\[-1.8ex]
		Label                & \texttt{pct5}\\
		$\nu_\minn$          & $-\infty$\\
		$p_i$                & 0.969 & 0.986 & 0.994 & 0.997 & 0.999\\
		$x_i$ for $\pi^\inp$ & 3.5   & 4.1   & 4.9   & 5.7   & 7.0\\
		\hline\\[-1.8ex]
		Label                & \texttt{cut5}\\
		$\nu_\minn$          & 3\\
		$p_i$                & 0.5   & 0.776 & 0.9   & 0.955 & 0.98\\
		$x_i$ for $\pi^\inp$ & 3.5   & 4.1   & 4.9   & 5.7   & 6.7\\
		\hline
	\end{tabular}
	\label{tab:observable_types}
\end{table}

Observable vectors are constructed by the description above with the settings of \tab{tab:observable_types}. This choice of bins and $p_i$ is made such that the same component from different types of observables represents about the same information, since the bin center of $\bx^{\abd 5}$  roughly correspond to $\bx^{\cut 5}$ for the input cosmology $\bpi^\inp$. Following \citetalias{Liu_etal_2015a}, who discovered in their study that the binwidth choice has a minor impact on parameter constraints if the estimated number count in each bin is $\gtrsim$ 10, we chose not to explore different choices of binwidths for $\bx^{\abd 5}$. We also note that $p_i$ for $x_i^{\cut 5}$ are logarithmically spaced.

By construction, the correlation between terms of percentile-like vectors is much higher than for the case of peak abundance. This tendency is shown in \tab{tab:corr_matrices} for the $\bpi^\inp$ cosmology. We discovered that $\bx^{\pct 5}$ and $\bx^{\cut 5}$ are highly correlated, while for $\bx^{\abd 5}$, the highest absolute value of off-diagonal terms does not exceed 17\%. A similar result was observed when we binned data differently. This suggests that the covariance should be included in likelihood analyses.

\begin{table}
	\centering
	\caption{Correlation matrices of $\bx^{\abd 5}$, $\bx^{\pct 5}$, and $\bx^{\cut 5}$ in the input cosmology. For $\bx^{\abd 5}$, the peak abundance is weakly correlated between bins.}
	\begin{tabular}{cc}
		\texttt{abd5} & 
		$\begin{pmatrix}
			1     & -0.05 & -0.09 & -0.08 & -0.16\\
			-0.05 & 1     & -0.05 & -0.01 & -0.12\\
			-0.09 & -0.05 & 1     & -0.04 & -0.11\\
			-0.08 & -0.01 & -0.04 & 1     & -0.06\\
			-0.16 & -0.12 & -0.11 & -0.06 & 1
		\end{pmatrix}$\\[7ex]
		\texttt{pct5} & 
		$\begin{pmatrix}
			1    & 0.62 & 0.29 & 0.15 & 0.11\\
			0.62 & 1    & 0.58 & 0.36 & 0.25\\
			0.29 & 0.58 & 1    & 0.66 & 0.43\\
			0.15 & 0.36 & 0.66 & 1    & 0.59\\
			0.11 & 0.25 & 0.43 & 0.59 & 1
		\end{pmatrix}$\\[7ex]
		\texttt{cut5} & 
		$\begin{pmatrix}
			1    & 0.58 & 0.31 & 0.20 & 0.15\\
			0.58 & 1    & 0.61 & 0.39 & 0.28\\
			0.31 & 0.61 & 1    & 0.65 & 0.47\\
			0.20 & 0.39 & 0.65 & 1    & 0.70\\
			0.15 & 0.28 & 0.47 & 0.70 & 1
		\end{pmatrix}$
	\end{tabular}
	\label{tab:corr_matrices}
\end{table}

\subsection{Constraint qualification}
\label{sect:methodology_qualification}

In this paper, both Bayesian inferences and likelihood-ratio tests \citep[see, e.g., Theorem 10.3.3 from][]{Casella_Berger_2002} have been performed. To distinguish between these two cases, we call \emph{credible region} the posterior PDF obtained from the Bayesian approach, which differs from the \emph{confidence region}, whose interpretation can be found in \sect{sect:nonParam_pValue}.

\begin{figure*}[tb]
	\centering
	\includegraphics[width=14cm]{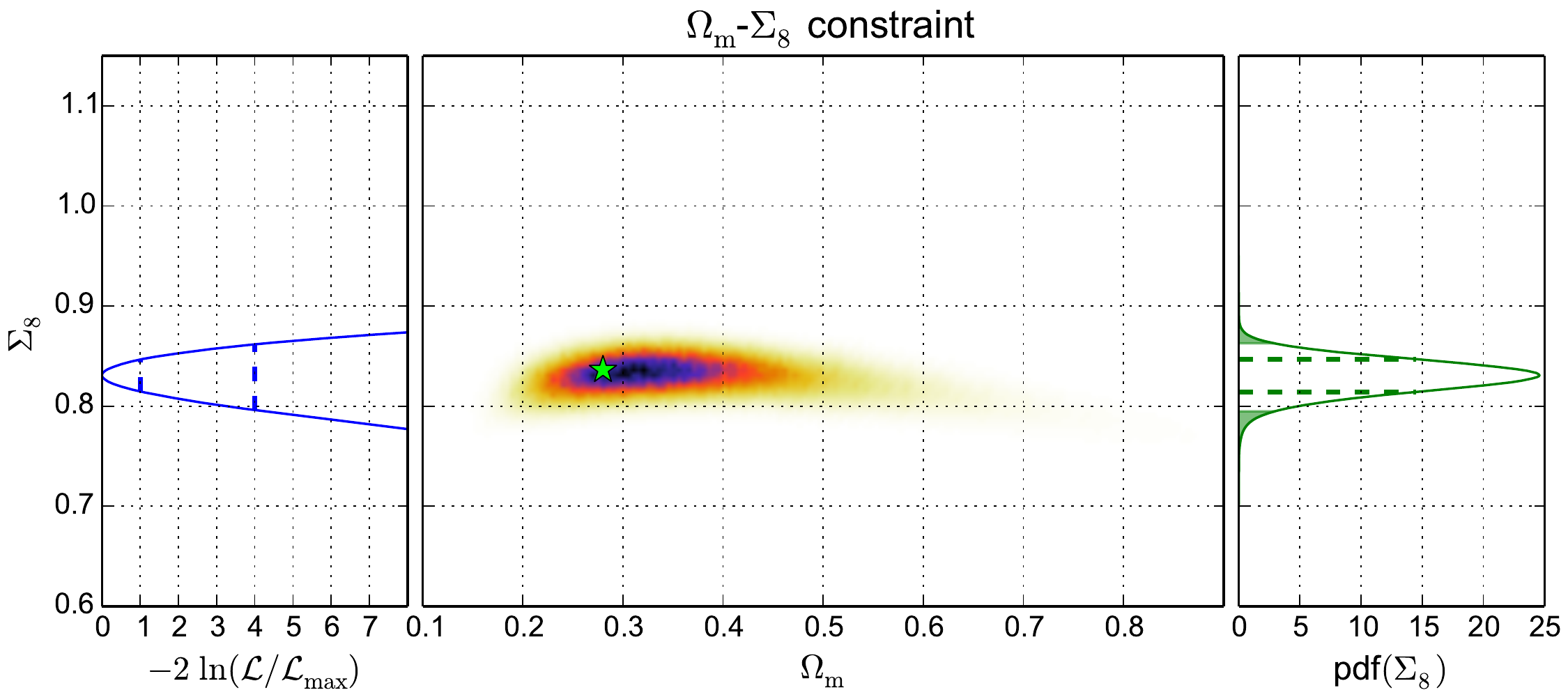}
	\caption{\textit{Middle panel}: the likelihood value using $\bx^{\abd 5}$ on the $\Omegam$-$\Sigma_8$ plane. The green star represents the input cosmology $\bpi^\inp$. Since $\log\sigeig$ and $\log\Omegam$ form an approximately linear degenerency, the quantity $\Sigma_8 \equiv \sigeig (\Omegam/0.27)^\alpha$ allows us to characterize the banana-shape contour thickness. \textit{Right panel}: the marginalized PDF of $\Sigma_8$. The dashed lines give the 1-$\sigma$ interval (68.3\%), while the borders of the shaded areas represent 2-$\sigma$ limits (95.4\%). \textit{Left panel:} the log-value of the marginalized likelihood ratio. Dashed lines in the \textit{left panel} give the corresponding value for 1 and 2-$\sigma$ significance levels, respectively.}
	\label{fig:pdf_Sigma8}
\end{figure*}

To quantify parameter-constraint contours, we introduce two criteria. Inspired by \cite{Jain_Seljak_1997} and \cite{Maoli_etal_2001}, the first criterion is to determine the error on
\begin{align}
	\Sigma_8 \equiv \sigeig(\Omegam/0.27)^\alpha.
  \label{for:Sigma_8}
\end{align}
Since the banana-shaped contour becomes more or less an elongated ellipse in log space, $\Delta\Sigma_8$ represents the ``thickness'' of the banana, tilted by the slope $\alpha$. Therefore, we first fit $\alpha$ with the linear relation $\log\Sigma_8 = \log\sigeig + \alpha\log(\Omegam/0.27)$, and then calculate the 1-$\sigma$ interval of $\Sigma_8$ on the $\Omegam$-$\Sigma_8$ plane. For a Bayesian approach, this interval is given by the 68\% most probable interval from the marginalized likelihood, while for a frequentist approach, significance levels are given by likelihood-ratio tests on the marginalized likelihood. Examples of both approaches are shown by \fig{fig:pdf_Sigma8}. Since no real data are used in this study, we are not interested in the best fit value, but the 1-$\sigma$ interval width $\Delta\Sigma_8$.

The second indicator is the figure of merit (FoM) for $\Omegam$ and $\sigma_8$, proposed by \cite{Dietrich_Hartlap_2010}. They define a FoM similar to the one from \citet{Albrecht_etal_2006} as the inverse of the area of the 2-$\sigma$ region.

\section{Influence of the cosmology-dependent covariance}
\label{sect:CDC}

\subsection{Formalism}
\label{sect:CDC_formalism}

In this section, we examine the cosmology-dependent-covariance (CDC) effect. From our statistic, we estimate the inverse covariance from \for{for:invCov_estimator} for each $\bpi$ from 1000 realizations.
By setting the Bayesian evidence to unity, $P(\bx=\bx^\obs)=1$, we write the relation among
prior probability $\mathcal{P}(\bpi)$, the likelihood $\mathcal{L}(\bpi)\equiv P(\bx^\obs|\bpi)$, 
and posterior probability $\mathcal{P}(\bpi|\bx^\obs)$ as
\begin{align}
	\mathcal{P}(\bpi|\bx^\obs) = \mathcal{L}(\bpi) \mathcal{P}(\bpi).
\end{align}
Given a model, we write $\Delta\bx(\bpi) \equiv
\bx^\model(\bpi)-\bx^\obs$ as the difference between the model prediction
$\bx^\model$ and the observation $\bx^\obs$. Then the Gaussian log-likelihood
is given by
\begin{align}
	-2\ln\mathcal{L}(\bpi) = \ln\left[(2\pi)^d\det(\bC) \right] 
	+ \Delta \bx^T  \bC\inv\Delta \bx
\end{align}
where
$d$ denotes the dimension of the observable space, and $\bC$ is the covariance
matrix for $\bx^\model$.

Estimating $\bC$ as an ensemble average is difficult since cosmologists only have one Universe. One can derive $\bC$ from observations with statistical techniques, such as bootstrap or jackknife \citepalias{Liu_etal_2015a}, or from a sufficient number of independent fields of view from $N$-body simulations \citepalias{Liu_etal_2015} or using analytic calculations. However, the first method only provides the estimation for a specific parameter set $\bC(\bpi = \bpi^\obs)$; the second method is limited to a small amount of parameters owing to the very high computational time cost; and the third method involves higher order statistics of the observables, which might not be well known. Thus, most studies suppose that the covariance matrix is invariant so ignore the CDC effect. In this case, the determinant term becomes a constant, and likelihood analysis can be summed up as the minimization of $\chi^2 \equiv \Delta \bx^T \bC\inv\Delta \bx$.

Alternatively, the stochastic characteristic of our model provides a quick and simple way to estimate the covariance matrix $\bC(\bpi)$ of each single parameter set $\bpi$. To examine the impact of the CDC effect in the peak-count framework, we write down the constant-covariance Gaussian (labeled \texttt{cg}), the semi-varying-covariance Gaussian (labeled \texttt{svg}), and the varying-covariance Gaussian (labeled \texttt{vg}) log-likelihoods as
\begin{align}
	L_\cg  &\equiv \Delta \bx^T(\bpi)\ \widehat{\bC\inv}(\bpi^\obs)\ \Delta \bx(\bpi), \label{for:L_cg}\\
	L_\svg &\equiv \Delta \bx^T(\bpi)\ \widehat{\bC\inv}(\bpi)\ \Delta \bx(\bpi),\ \ \text{and}\\
	L_\vg &\equiv \ln\left[\det \widehat{\bC}(\bpi)\right] + \Delta\bx^T(\bpi)\ \widehat{\bC\inv}(\bpi)\ \Delta \bx(\bpi).
\end{align}
Here, the term $\widehat{\bC\inv}(\bpi^\obs)$ in \for{for:L_cg} refers to $\widehat{\bC\inv}(\bpi^\inp)$, where $\bpi^\inp$ is described in \sect{sect:methodology_design}. By comparing the contours derived from different likelihoods, we aim to measure (1) the evolution of the $\chi^2$ term by substituting the constant matrix with the true varying $\widehat{\bC\inv}$, and (2) the impact from adding the determinant term. Therefore, $L_\svg$ is just an illustrative case to assess the influence of the two terms in the likelihood.

\subsection{The $\chi^2$ term}
\label{sect:CDC_chi2}

\begin{figure*}[tb]
	\centering
	\includegraphics[width=8.5cm]{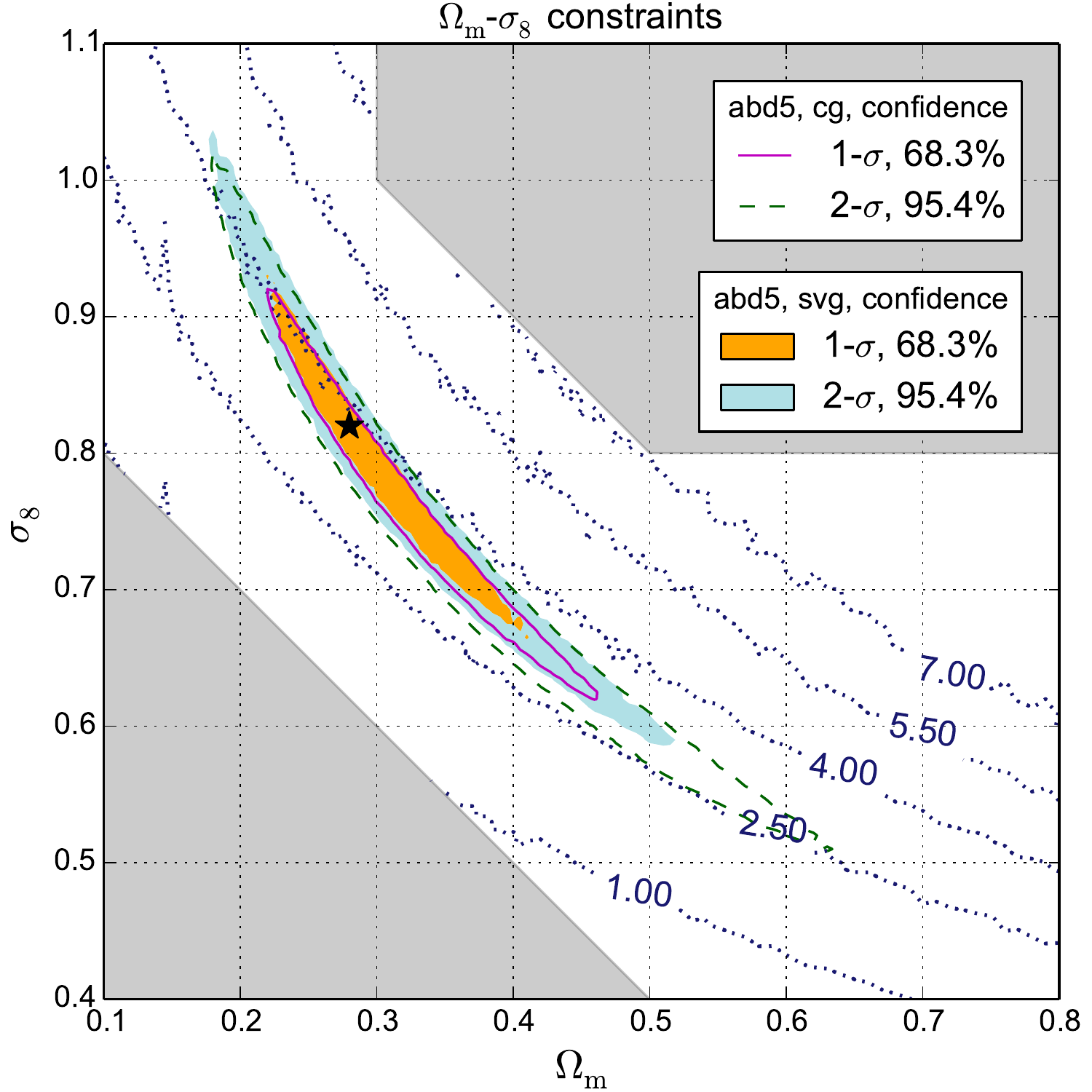}
	\includegraphics[width=8.5cm]{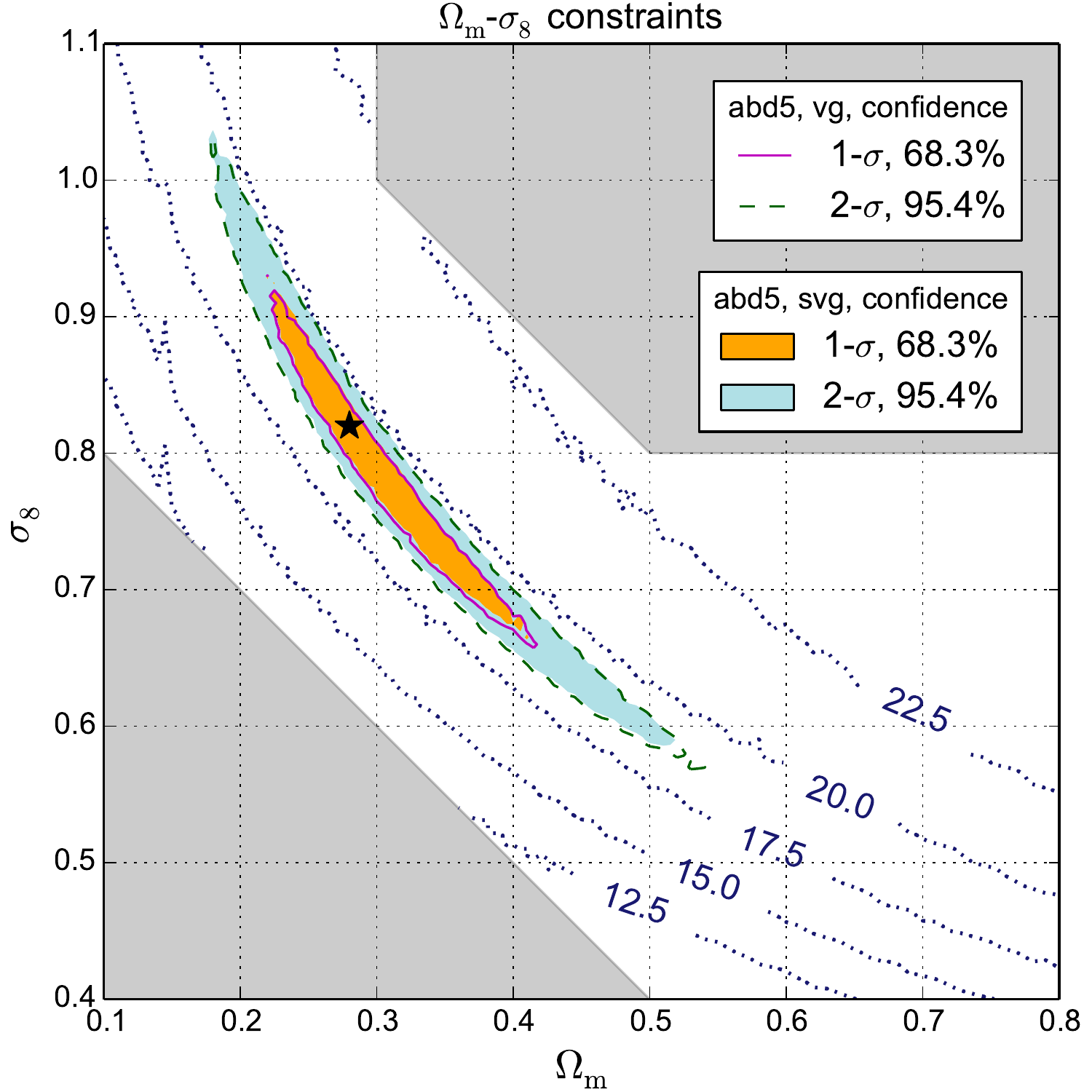}
	\caption{Confidence regions derived from $L_\cg$, $L_\svg$, and $L_\vg$ with $\bx^{\abd 5}$. The solid and dashed lines represent $L_\cg$ in the \textit{left panel} and $L_\vg$ in the \textit{right panel}, while the colored areas are from $L_\svg$. The black star stands for $\bpi^\inp$ and gray areas represent the non-explored parameter space. The dotted lines are different isolines, the variance $\hat{C}_{55}$ of the bin with highest S/N in the \textit{left panel} and $\ln(\det\hat{C})$ for the \textit{right panel}. The contour area is reduced by 22\% when taking the CDC effect into account. The parameter-dependent determinant term does not contribute significantly.}
	\label{fig:contour_Gaussian_abd5}
\end{figure*}

\begin{figure*}[tb]
	\includegraphics[width=8.5cm]{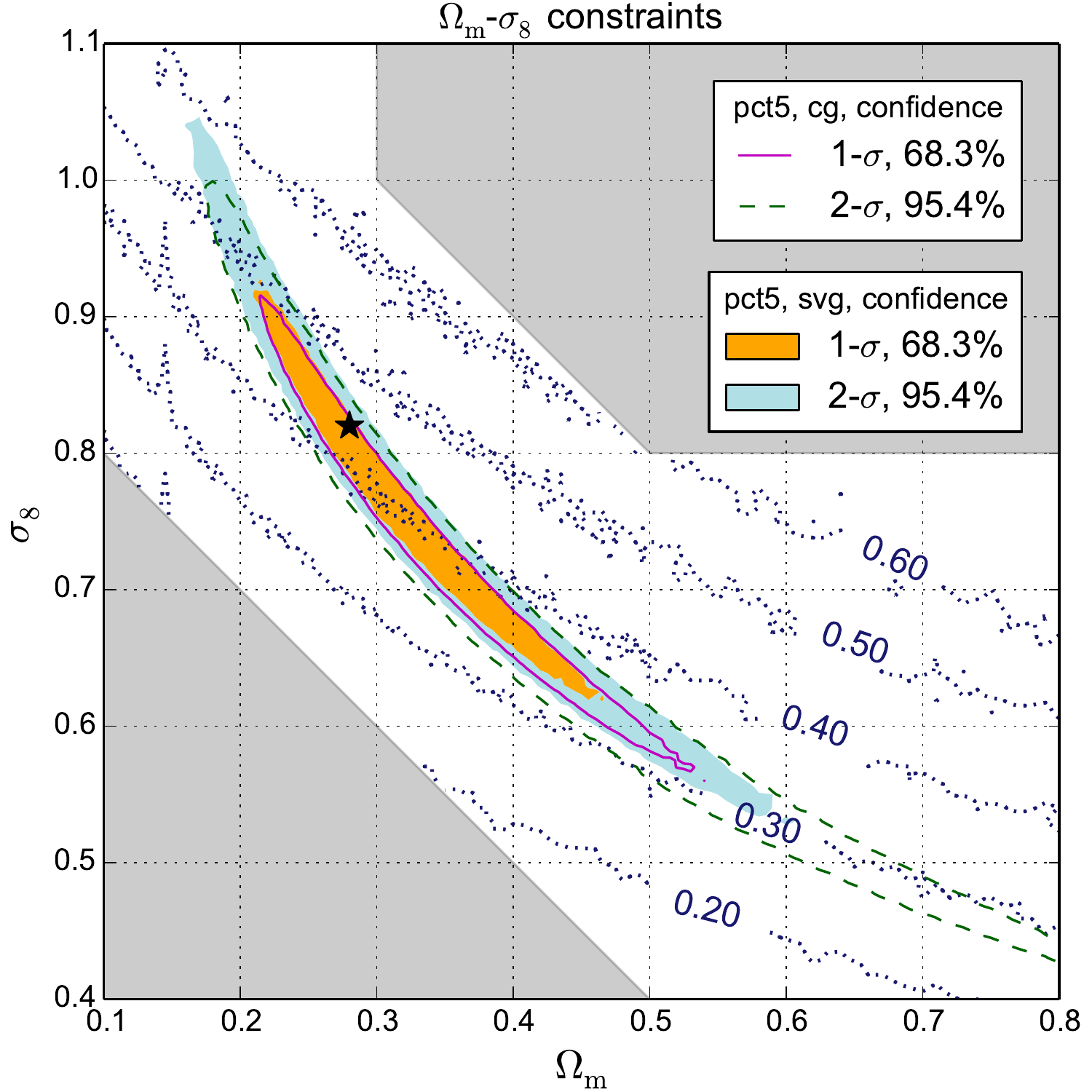}
	\includegraphics[width=8.5cm]{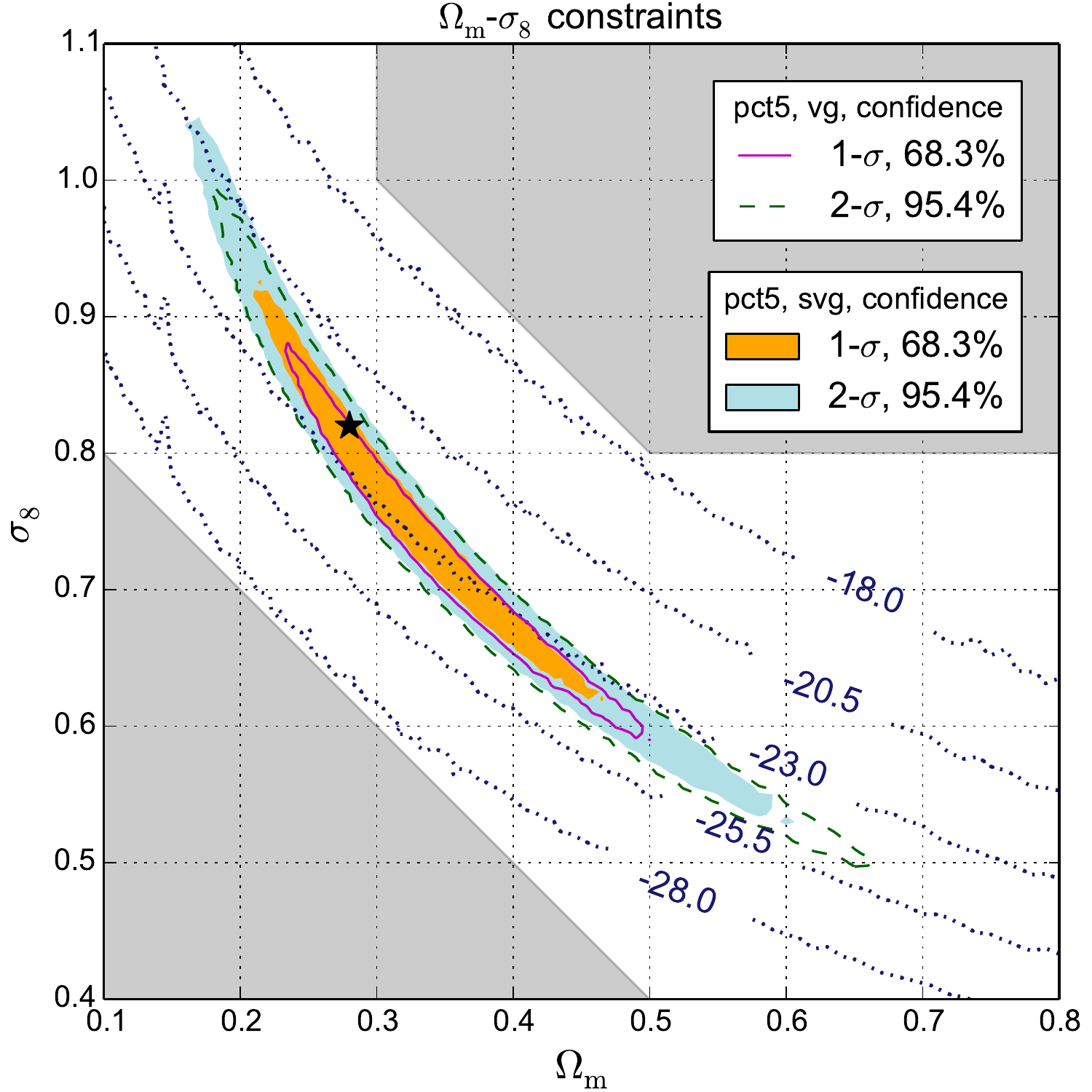}\\
	\includegraphics[width=8.5cm]{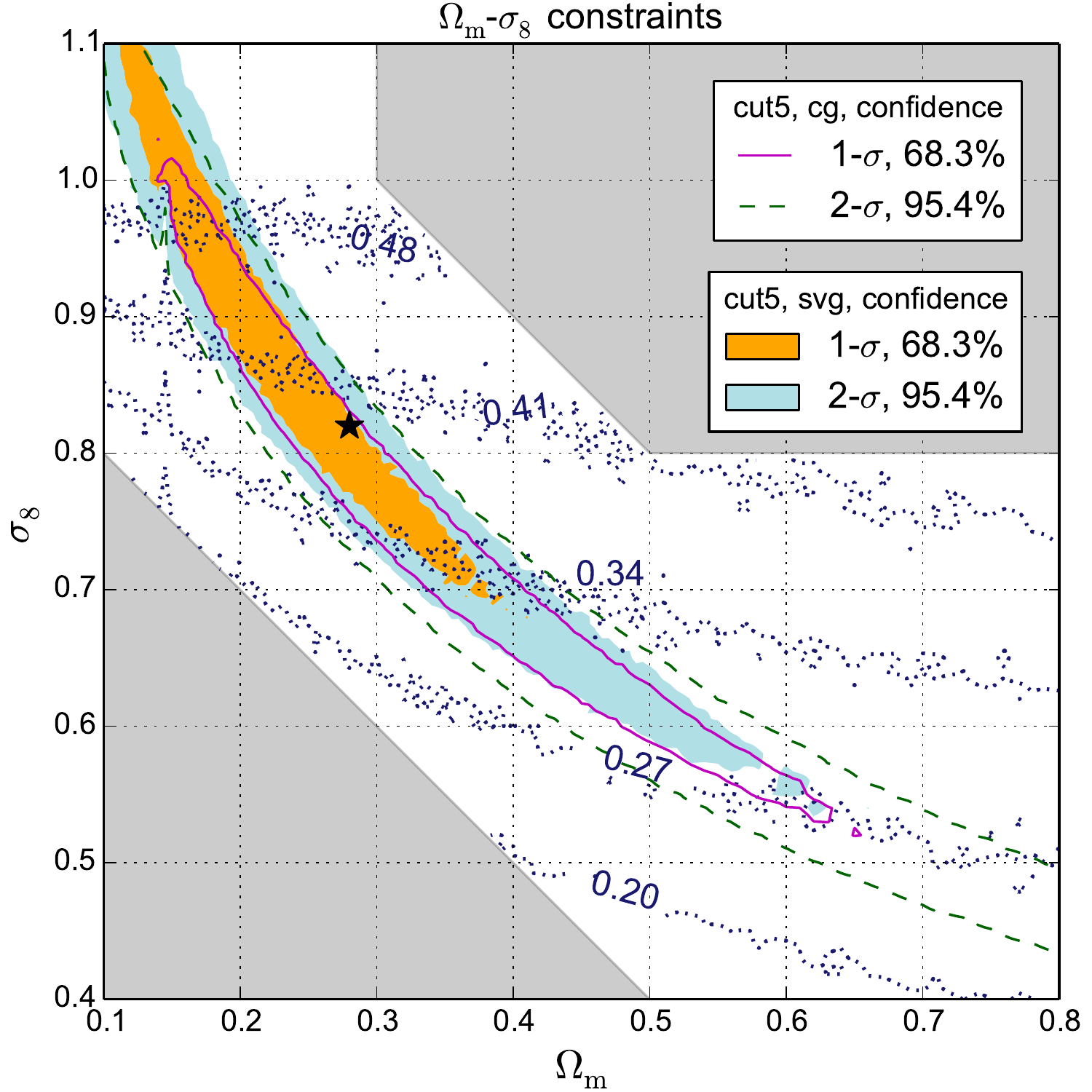}
	\includegraphics[width=8.5cm]{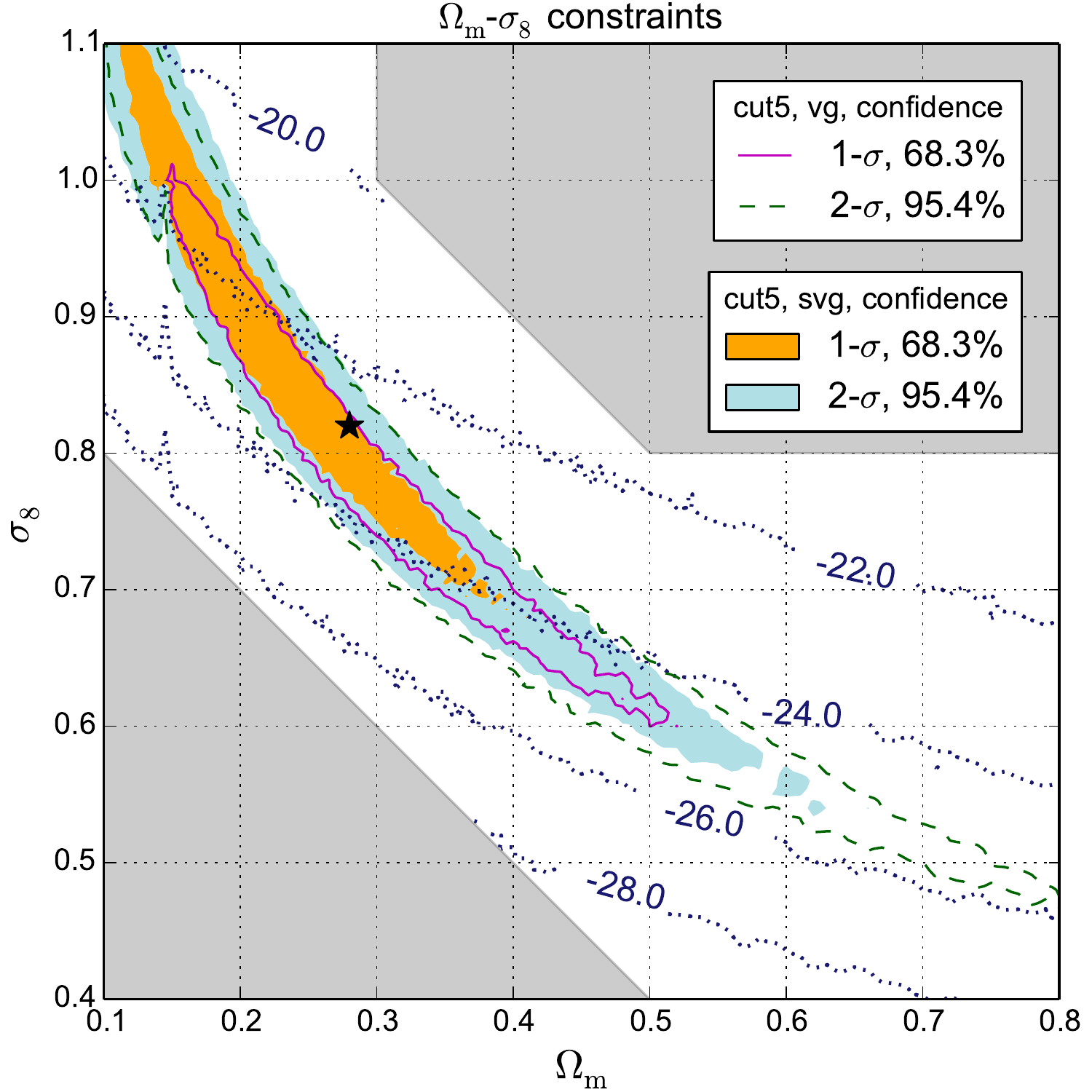}
    \caption{Similar to \fig{fig:contour_Gaussian_abd5}. Confidence regions with $\bx^{\pct 5}$ and $\bx^{\cut 5}$. Both \textit{upper panels} are drawn with $\bx^{\pct 5}$ and both \textit{lower panels} with $\bx^{\cut 5}$. Both \textit{left panels} are the comparison between $L_\cg$ and $L_\svg$, and both \textit{right panels} between $L_\svg$ and $L_\vg$.}
	\label{fig:contour_Gaussian_pct5}
\end{figure*}

The lefthand panel of \fig{fig:contour_Gaussian_abd5} shows the comparison between confidence regions derived from $L_\cg$ and $L_\svg$ with $\bx^\abd$. It shows a clear difference of the contours between $L_\cg$ and $L_\svg$. Since the off-diagonal correlation coefficients are weak (as shown in \tab{tab:corr_matrices}), the variation in diagonal terms of $\bC$ plays a major role in the size of credible regions. The isolines for $\hat{C}_{55}$ are also drawn in \fig{fig:contour_Gaussian_abd5}. These isolines cross the $\Omegam$-$\sigeig$ degenerency lines from $L_\cg$ and thus shrink the credible region. We also find that the isolines for $\hat{C}_{11}$ and $\hat{C}_{22}$ are noisy and that those for $\hat{C}_{33}$ and $\hat{C}_{44}$ coincide well with the original degeneracy direction.

\tab{tab:criteria_confidence} shows the values of both criteria for different likelihoods. We observe that using $L_\svg$ significantly improves the constraints by 24\% in terms of FoM. Regarding $\Delta\Sigma_8$, the improvement is weak. As a result, using varying covariance matrices breaks down part of the banana-shape degenerency and shrinks the contour length, but does not reduce the thickness.

In the lefthand panels of \fig{fig:contour_Gaussian_pct5}, we show the same
constraints derived from two other observables $\bx^{\pct 5}$ and $\bx^{\cut
5}$. We see a similar CDC effect for both. We observe that $\bx^{\pct 5}$ has
less constraining power than $\bx^{\abd 5}$, and $\bx^{\cut 5}$ is
outperformed by both other data vectors. This is due to the cutoff value
$\nu_\minn$. Introducing a cutoff at $\nu_\minn = 3$ decreases the total number
of peaks and amplifies the fluctuation of high-peak values in the CDF. When we
use percentiles to define observables, the distribution of each component of
$\bx^{\cut 5}$ becomes wider than the one of the corresponding component of
$\bx^{\pct 5}$, and this greater scatter in the CDF enlarges the contours.
However, the cutoff also introduces a tilt for the contours.
\tab{tab:best_fit_confidence} shows the best fit $\alpha$ for the different
cases. The difference in the tilt could be a useful tool for improving the
constraining power. This has also been observed by
\cite{Dietrich_Hartlap_2010}. Nevertheless, we do not take on any joint
analysis since $\bx^{\abd 5}$ and $\bx^{\cut 5}$ contain essentially the same
information.

\subsection{Impact from the determinant term}
\label{sect:CDC_determinant}

The righthand panel of \fig{fig:contour_Gaussian_abd5} shows the comparison between $L_\svg$ and $L_\vg$ with $\bx^{\abd 5}$. It shows that adding the determinant term does not result in significant changes of the parameter constraints. The isolines from $\ln(\det\hat{C})$ explain this, since the gradients are perpendicular to the degenerency lines. We observe that including the determinant makes the contours slightly larger, but almost negligibly so. The total improvement in the contour area compared to $L_\cg$ is 22\%.

However, a different change is seen for $\bx^{\pct 5}$ and $\bx^{\cut 5}$. Adding the determinant to the likelihood computed from these observables induces a shift of contours toward the higher $\Omegam$ area. In the case of $\bx^{\cut 5}$, this shift compensates for the contour offset from the varying $\chi^2$ term, but does not improve either $\Delta\Sigma_8$ or FoM significantly, as shown in \tab{tab:criteria_confidence}. As a result, using the Gaussian likelihood, the total CDC effect can be summed up as an improvement of at least 14\% in terms of thickness and 38\% in terms of area.

The results from Bayesian inference is very similar to the likelihood-ratio test. Thus, we only show their $\Delta\Sigma_8$ and FoM in \tab{tab:criteria_credible} and best fits in \tab{tab:best_fit_credible}. We recall that a similar analysis was done by \cite{Eifler_etal_2009} on shear covariances. Our observations agree with their conclusions: a relatively large impact from the $\chi^2$ term and negligible change from the determinant term. However, the total CDC effect is more significant in the peak-count framework than for the power spectrum.

\section{Testing the copula transform}
\label{sect:copula}

\subsection{Formalism}
\label{sect:copula_formalism}

Consider a multivariate joint distribution $P(x_1, \ldots, x_d)$. In general, $P$ could be far from Gaussian so that imposing a Gaussian likelihood could induce biases. The idea of the copula technique is to evaluate the likelihood in a new observable space where the Gaussian approximation is better. Using a change in variables, individual marginalized distributions of $P$ can be approximated to Gaussian ones. This is achieved by a series of successive one-dimensional, axis-wise transformations. The multivariate Gaussianity of the transformed distribution is \textbf{not} garanteed. However, in some cases, this transformation tunes the distribution and makes it more ``Gaussian'', so that evaluating the likelihood in the tuned space is more realistic \citep{Benabed_etal_2009, Sato_etal_2011}.

From Sklar's theorem \citep{Sklar_1959}, any multivariate distribution $P(x_1, \ldots, x_d)$ can be decomposed into the copula density multiplied by marginalized distributions. A comprehensible and elegant demonstration is given by \cite{Rueschendorf_2009}. Readers are also encouraged to follow \cite{Scherrer_etal_2010} for detailed physical interpretations and \cite{Sato_etal_2011} for a very pedagogical derivation of the Gaussian copula transform.

Consider a $d$-dimensional distribution $P(\bx)$, where $\bx = (x_1, \ldots, x_d)$ is a random vector. We let $P_i$ be the marginalized 1-point PDF of $x_i$, and $F_i$ the corresponding CDF. Sklar's theorem shows that there is a unique $d$-dimensional function $c$ defined on $[0, 1]^d$ with uniform marginal PDF, such that
\begin{align}\label{for:copula_relation}
	P(\bx) = c(\vect{u}) P_1(x_1) \cdots P_d(x_d),
\end{align}
where $u_i \equiv F_i(x_i)$. The function $c$ is called the \emph{copula density}. On the other hand, let $q_i \equiv \Phi_i\inv(u_i)$, where $\Phi_i$ is the CDF of the normal distribution with the same means $\mu_i$ and variances $\sigma_i^2$ as the laws $P_i$, such that
\begin{align}
	\Phi_i(q_i) &\equiv \int_{-\infty}^{q_i} \phi_i(q') \rmd q',\\
	\phi_i(q_i) &\equiv \frac{1}{\sqrt{2\pi\sigma_i^2}} \ \exp\left[ -\frac{(q_i-\mu_i)^2}{2\sigma^2_i} \right].
\end{align}
We can then define a new joint PDF $P'$ in the $\vect{q}$ space that corresponds to $P$ in $\bx$ space, i.e. $P'(\vect{q}) = P(\bx)$. The marginal PDF and CDF of $P'$ are only $\phi_i$ and $\Phi_i$, respectively. Thus, applying \for{for:copula_relation} to $P'$ and $\phi_i$ leads to
\begin{align}\label{for:copula_relation_2}
	P'(\vect{q}) = c(\vect{u}) \phi_1(q_1) \cdots \phi_d(q_d).
\end{align}
By the uniqueness of the copula density, $c$ in Eqs. \eqref{for:copula_relation} and \eqref{for:copula_relation_2} are the same. Thus, we obtain
\begin{align}
	P(\bx) = P'(\vect{q}) \frac{P_1(x_1) \cdots P_d(x_d)}{\phi_1(q_1) \cdots \phi_d(q_d)}.
\end{align}
We note that the marginal PDFs of $P'$ are identical to a multivariate Gaussian distribution $\phi$ with mean $\vect{\mu}$ and covariance $\bC$, where $\bC$ is the covariance matrix of $\bx$. The PDF of $\phi$ is given by
\begin{align}
	\phi(\vect{q}) &\equiv \frac{1}{\sqrt{(2\pi)^d\det\bC}} \exp\left[ -\frac{1}{2}\sum_{i,j}(q_i-\mu_i) C_{ij}\inv (q_j-\mu_j)\right].
\end{align}
Finally, by approximating $P'$ to $\phi$, one gets the \emph{Gaussian copula transform}:
\begin{align}\label{for:copula_transform}
	P(\bx) = \phi(\vect{q}) \frac{P_1(x_1) \cdots P_d(x_d)}{\phi_1(q_1) \cdots \phi_d(q_d)}.
\end{align}

Why is it more accurate to calculate the likelihood in this way? In the classical case, since the shape of $P(\bx)$ is unknown, we approximate it to a normal distribution: $P(\bx) \approx \phi(\bx)$. Applying the Gaussian copula transform means that we carry out this approximation in the new space of $\vect{q}$: $P'(\vect{q}) \approx \phi(\vect{q})$. Since $q_i = \Phi\inv_i(F_i(x_i))$, at least the marginals of $P'(\vect{q})$ are strictly Gaussian. And \for{for:copula_transform} gives the corresponding value in $\bx$ space, while taking $P'(\vect{q}) \approx \phi(\vect{q})$ in $\vect{q}$ space. However, in some cases, the copula has no effect at all. We consider $f(x,y) = 2\phi_2(x,y)\Theta(xy)$ where $\phi_2$ is the two-dimensional standard normal distribution, and $\Theta$ is the Heaviside step function. The value of $f$ is two times $\phi_2$ if $x$ and $y$ have the same sign and 0 otherwise. The marginal PDF of $f$ and $\phi_2$ turn out to be the same. As a result, the Gaussian copula transform does nothing and $f$ remains extremely non-Gaussian. However, if we do not have any prior knowledge, then the result with the copula transformation should be at least as good as the classical likelihood.

By applying \for{for:copula_transform} to $P(\bx^\obs|\bpi)$, one gets the \emph{copula likelihood}:
\begin{align}
	\mathcal{L}(\bpi) &= \frac{1}{\sqrt{(2\pi)^d\det\bC}} \notag\\
	&\times \exp\left[ -\frac{1}{2}\sum_{i=1}^d\sum_{j=1}^d \left(q^\obs_i-\mu_i\right) C_{ij}\inv \left(q^\obs_j-\mu_j\right)\right] \notag\\
	&\times \prod_{i=1}^d \left[\frac{1}{\sqrt{2\pi\sigma_i^2}} \ \exp\left( -\frac{\left(q^\obs_i-\mu_i\right)^2}{2\sigma^2_i} \right)\right]\inv \prod_{i=1}^d P_i(x_i^\obs). \label{for:copula_likelihood}
\end{align}
In this paper, $\mu_i = x_i^\model$. Including the dependency on $\bpi$ for all relevant quantities, the varying-covariance copula log-likelihood $L_\vc$ is given by
\begin{align}
	L_\vc &\equiv \ln\left[\det \widehat{\bC}(\bpi)\right] \notag\\
	&+ \sum_{i=1}^d \sum_{j=1}^d \left(q_i^\obs(\bpi) - x^\model_i(\bpi)\right) \widehat{C\inv_{ij}}(\bpi) \left(q_j^\obs(\bpi) - x^\model_j(\bpi)\right) \notag\\
	&- 2 \sum_{i=1}^d \ln\hat{\sigma}_i(\bpi) - \sum_{i=1}^d \left(\frac{q_i^\obs(\bpi) - x^\model_i(\bpi)}{\hat{\sigma}_i(\bpi)}\right)^2 \notag\\
	&- 2 \sum_{i=1}^d \ln\hat{P}_i(x^\obs_i|\bpi).\label{for:L_vc}
\end{align}
Here, $\hat{P}_i(\cdot|\bpi)$ is the $i$-th marginal $\bpi$-dependent PDF that we estimate directly from the $N$ samples $x^{(k)}_i, k=1\ldots N$ already mentioned in \sect{sect:methodology_design}, using the kernel density estimation (KDE):
\begin{align}
	\hat{P}_i(x_i) = \frac{1}{N} \sum_{k=1}^N \frac{1}{h_i}K\left( \frac{x_i - x_i\upp{k}}{h_i} \right),
\end{align}
where the kernel $K$ is Gaussian, and the bandwidth $h_i = (4/3N)^{1/5}\hat{\sigma}_i$ is given by Silverman's rule \citep{Silverman_1986}. These are one-dimensional PDF estimations, and the time cost is almost negligible. The term $\hat{P}_i(x^\obs_i|\bpi)$ should be understood as a one-point evaluation of this function at $x^\obs_i$. The quantities $x^\model_i(\bpi)$, $\hat{\sigma}_i(\bpi)$, and $\widehat{C\inv_{ij}}(\bpi)$ are estimated with the same set following Eqs. \eqref{for:mean_estimator}, \eqref{for:cov_estimator}, and \eqref{for:invCov_estimator}. Finally, $q^\obs_i(\bpi) = \Phi_i\inv(\hat{F}_i(x^\obs_i|\bpi))$. We highlight that $\hat{F}_i(\cdot|\bpi)$ is the CDF that corresponds to $\hat{P}_i(\cdot|\bpi)$, and $\Phi_i$ also depends on $\bpi$ via $\mu_i$ and $\hat{\sigma}_i$.

We are also interested in studying the copula transform under the constant-covariance situation. In this case, we define the constant-covariance copula log likelihood $L_\cc$ as
\begin{align}
	L_\cc &\equiv \sum_{i=1}^d \sum_{j=1}^d \left(q_i^\obs(\bpi) - x^\model_i(\bpi)\right) \widehat{C\inv_{ij}}(\bpi^\inp) \left(q_j^\obs(\bpi) - x^\model_j(\bpi)\right) \notag\\
	&- \sum_{i=1}^d \left(\frac{q_i^\obs(\bpi) - x^\model_i(\bpi)}{\hat{\sigma}_i(\bpi^\inp)}\right)^2 - 2 \sum_{i=1}^d \ln \hat{P}_i\left(x^\obs_i - x^\model_i(\bpi)\right). \label{for:L_cc}
\end{align}
Besides the constant covariance, we also suppose that the distribution of each $x_i$ around its mean value does not vary with $\bpi$. In this case, $\hat{P}_i(\cdot)$ denotes the zero-mean marginal PDF, and it is only estimated once from the 1000 realizations of $\bpi^\inp$, as are $\hat{\sigma}_i$ and $\widehat{C\inv_{ij}}$. We recall that $q^\obs_i(\bpi) = \Phi_i\inv(\hat{F}_i(x^\obs_i - x^\model_i(\bpi)))$ where $\Phi_i$ depends on $\bpi$ implicitly via $\mu_i$ and $\hat{\sigma}_i$.

\subsection{Constraints using the copula}
\label{sect:copula_constraints}

\begin{figure*}[tb]
	\centering
	\includegraphics[width=8.5cm]{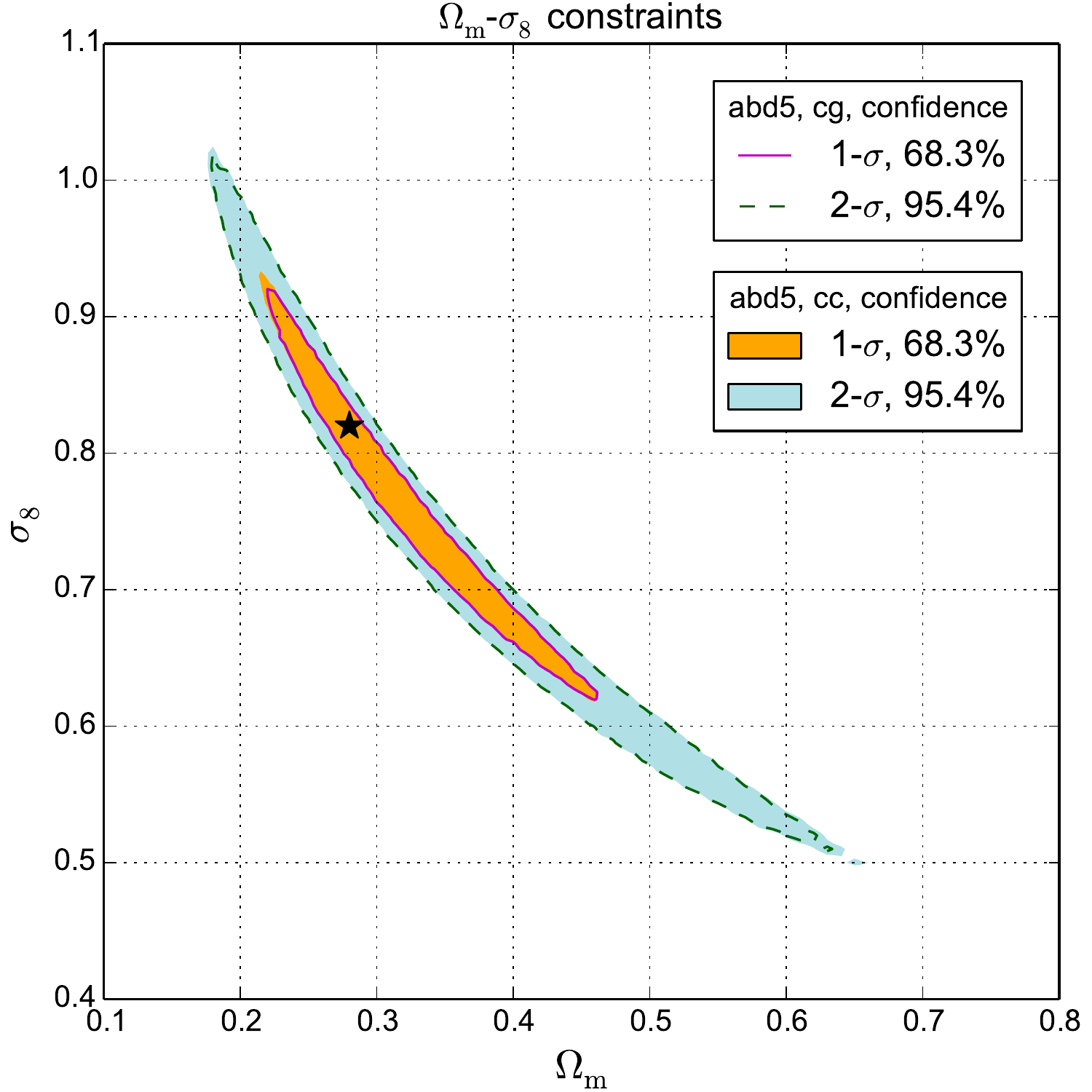}
	\includegraphics[width=8.5cm]{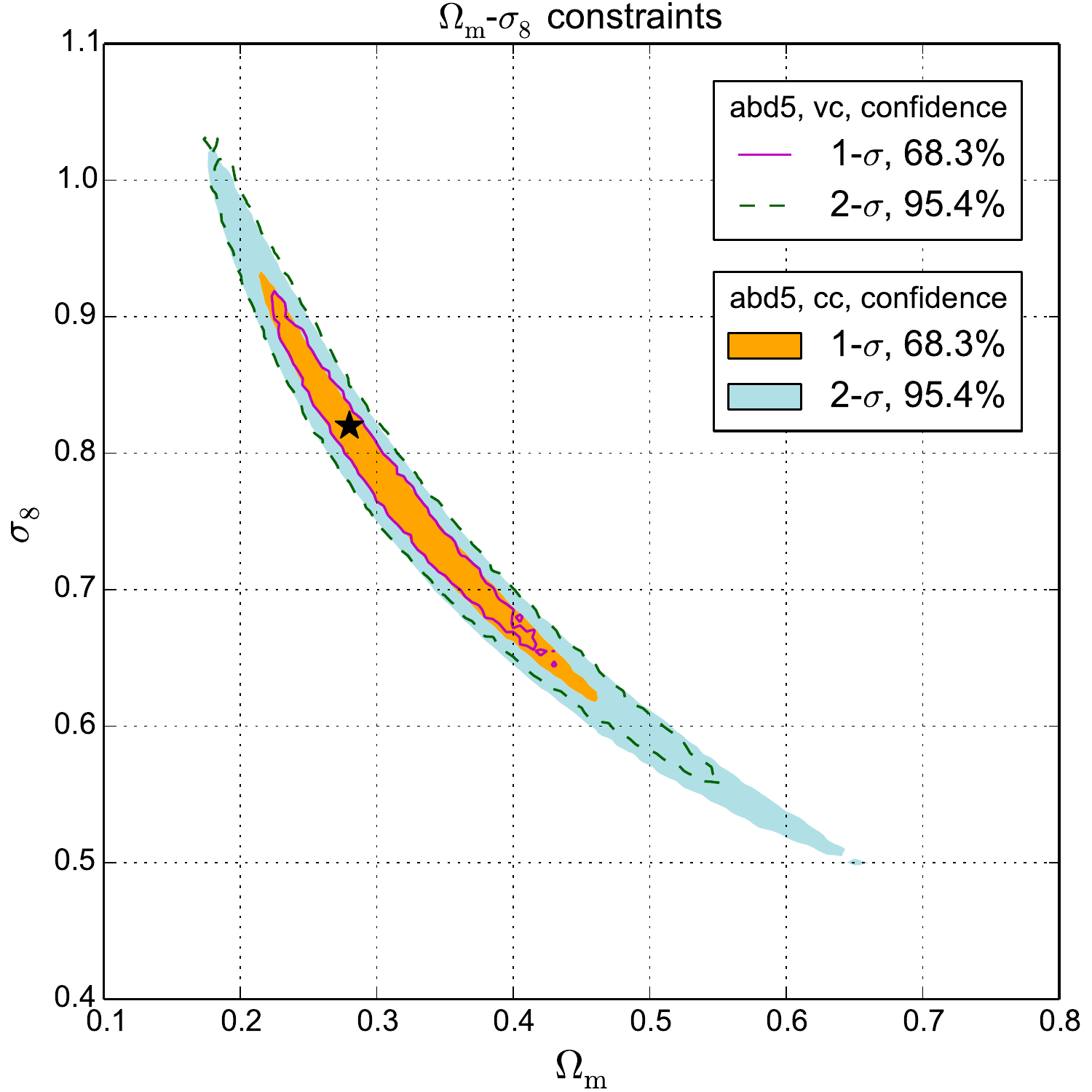}
	\caption{Confidence regions derived from copula analyses. \textit{Left panel}: comparison between contours from $L_\cg$ (solid and dashed lines) and $L_\cc$ (colored areas). \textit{Right panel}: comparison between contours from $L_\cc$ (colored areas) and $L_\vc$ (solid and dashed lines). The evolution tendency from $L_\cc$ to $L_\vc$ is similar to the evolution from $L_\cg$ to $L_\vg$.}
	\label{fig:contour_copula_abd5}
\end{figure*}

We again use the setting described in \sect{sect:methodology_design}. 
We outline two interesting comparisons, which are
shown in \fig{fig:contour_copula_abd5}: between $L_\cg$ and $L_\cc$ in the lefthand
panel and between $L_\cc$ and $L_\vc$ in the righthand one, both with $\bx^{\abd
5}$. The lefthand panel shows that, for weak-lensing peak counts, the Gaussian
likelihood is a very good approximation. Quantitative results, shown in Tables
\ref{tab:criteria_confidence}, \ref{tab:best_fit_confidence},
\ref{tab:criteria_credible}, and \ref{tab:best_fit_credible}, reveal that the
Gaussian likelihood provides slightly optimistic $\Omegam$-$\sigeig$ constraints.
We would like to emphasize that the effect of
the copula transform is ambiguous, and both tighter or wider constraints are
possible. This has already shown by \cite{Sato_etal_2011}, who found that the
Gaussian likelihood underestimates the constraint power for low $\ell$ of the
lensing power spectrum and overestimates it for high $\ell$.

In the righthand panel of \fig{fig:contour_copula_abd5}, when the CDC effect is
taken into account for the copula transform, the parameter constrains are
submitted to a similar change to the Gaussian likelihood. Tighter
constraints are obtained from $L_\vc$ than from $L_\cc$. Similar results can be
found for $\bx^{\pct 5}$ and $\bx^{\cut 5}$. In summary, the copula with
varying covariance, $L_\vc$ results in an FoM improvement of at least 10\%
compared to the Gaussian case with constant covariance, $L_\cg$.

\section{Non-analytic likelihood analyses}
\label{sect:nonParam}

\subsection{The true likelihood}
\label{sect:nonParam_likelihood}

In this section, we obtain the parameter constraints in a more direct way. Since our model predictions sample the full PDF, the PDF-Gaussianity assumption is no longer necessary. This allows us to go back to the true definition of the log-likelihood:
\begin{align}
	L_\true \equiv -2 \ln \hat{P}(\bx^\obs|\bpi),
\end{align}
where $\hat{P}$ is estimated from our $N$ realizations $\bx\upp{k}$ ($\bpi$-dependent) using the kernel density estimation technique. The multivariate estimation is performed by
\begin{align}\label{for:multivariate_KDE}
	\hat{P}(\bx) &= \frac{1}{N} \sum_{k=1}^N K(\bx-\bx\upp{k}),\\
	K(\bx)&= \frac{1}{\sqrt{(2\pi)^d|\det\vect{H}|}}\exp\left[-\frac{1}{2} \bx^T \vect{H}\inv \bx\right],
\end{align}
where
\begin{align}\label{for:multivariate_KDE_bandwidth}
	\sqrt{\vect{H}_{ij}} = \left\{\begin{matrix}
		\displaystyle
		\left[\frac{4}{(d+2)N}\right]^{\textstyle \frac{1}{d+4}}\hat{\sigma}_i & \text{if}\ \ i = j\ ,\\[3ex]
		0 & \text{otherwise.}
	\end{matrix}\right.
\end{align}
The evaluation of this non-analytic likelihood gets very noisy when the observable dimension $d$ increases. In this case, a larger $N$ will be required to stabilize the constraints. As in previous sections, we perform both the likelihood-ratio test and Bayesian inference with this likelihood.

\begin{table}
	\centering
    \caption{$\Delta\Sigma_8$, the error on the parameter (\ref{for:Sigma_8}) and the figure of merit (FoM) for confidence regions are summarized for the different analysis approaches performed in this paper. $L_\cg$, $L_\svg$, and $L_\vg$ are introduced in \sect{sect:CDC}, $L_\cc$ and $L_\vc$ in \sect{sect:copula}, and $L_\true$ and $p$-value in \sect{sect:nonParam}. In each case, we take $\bx^{\abd 5}$, $\bx^{\pct 5}$, or $\bx^{\cut 5}$ as data vector as indicated in the table rows.}
	\renewcommand{\arraystretch}{1.3}
	\begin{tabular}{c|cc|cc|cc}
		\hline\hline
		          & \multicolumn{2}{c|}{$\bx^{\abd 5}$} & \multicolumn{2}{c|}{$\bx^{\pct 5}$} & \multicolumn{2}{c}{$\bx^{\cut 5}$}\\[-0.5ex]
		\hline
		          & $\Delta\Sigma_8$ & FoM & $\Delta\Sigma_8$ & FoM & $\Delta\Sigma_8$ & FoM\\[-0.3ex]
		\hline
		$L_\cg$   & 0.032            & 46  & 0.037            & 31  & 0.065            & 13 \\[-0.5ex]
		$L_\svg$  & 0.031            & 57  & 0.032            & 42  & 0.054            & 21 \\[-0.5ex]
		$L_\vg$   & 0.031            & 56  & 0.032            & 43  & 0.052            & 18 \\[-0.5ex]
		$L_\cc$   & 0.032            & 43  & 0.038            & 33  & 0.056            & 13 \\[-0.5ex]
		$L_\vc$   & 0.033            & 52  & 0.034            & 39  & 0.058            & 16 \\[-0.5ex]
		$L_\true$ & 0.033            & 54  & 0.035            & 39  & 0.058            & 17 \\[-0.5ex]
		$p$-value & 0.035            & 39  & 0.037            & 27  & 0.067            & 12 \\
		\hline
	\end{tabular}
	\renewcommand{\arraystretch}{1.0}
	\label{tab:criteria_confidence}
\end{table}

\begin{table*}
	\centering
	\caption{Best fits of $(\Sigma_8, \alpha)$ from all analyses using the likelihood-ratio test and $p$-value analysis (confidence region). The description of $L_\cg$, $L_\svg$, and $L_\vg$ can be found in \sect{sect:CDC}, $L_\cc$ and $L_\cg$ in \sect{sect:copula}, and $L_\true$ and $p$-value in \sect{sect:nonParam}. We note that the best fits for $\Sigma_8$ are indicative since we do not use the real observational data in this study.}
	\renewcommand{\arraystretch}{1.3}
	\begin{tabular}{c|cc|cc|cc}
		\hline\hline
		              & \multicolumn{2}{c|}{$\bx^{\abd 5}$} & \multicolumn{2}{c|}{$\bx^{\pct 5}$} & \multicolumn{2}{c}{$\bx^{\cut 5}$}\\[-0.5ex]
		\hline
		              & $\Sigma_8$ $^{+1\sigma}_{-1\sigma}$ & $\alpha$ & $\Sigma_8$ $^{+1\sigma}_{-1\sigma}$ & $\alpha$ & $\Sigma_8$ $^{+1\sigma}_{-1\sigma}$ & $\alpha$\\[0.1ex]
		\hline
		$L_\cg$   & 0.831$^{+0.016}_{-0.016}$ & 0.54 & 0.822$^{+0.018}_{-0.019}$ & 0.54 & 0.800$^{+0.030}_{-0.035}$ & 0.45\\[0.2ex]
		$L_\svg$  & 0.831$^{+0.016}_{-0.015}$ & 0.52 & 0.820$^{+0.015}_{-0.016}$ & 0.51 & 0.800$^{+0.031}_{-0.023}$ & 0.40\\[0.2ex]
		$L_\vg$   & 0.829$^{+0.015}_{-0.015}$ & 0.52 & 0.819$^{+0.015}_{-0.016}$ & 0.52 & 0.800$^{+0.024}_{-0.028}$ & 0.42\\[0.2ex]
		$L_\cc$   & 0.830$^{+0.016}_{-0.016}$ & 0.54 & 0.825$^{+0.018}_{-0.020}$ & 0.54 & 0.807$^{+0.025}_{-0.031}$ & 0.46\\[0.2ex]
		$L_\vc$   & 0.829$^{+0.016}_{-0.016}$ & 0.52 & 0.823$^{+0.016}_{-0.019}$ & 0.53 & 0.798$^{+0.029}_{-0.029}$ & 0.44\\[0.2ex]
		$L_\true$ & 0.828$^{+0.018}_{-0.015}$ & 0.53 & 0.823$^{+0.015}_{-0.020}$ & 0.53 & 0.800$^{+0.028}_{-0.030}$ & 0.44\\[0.2ex]
		$p$-value & 0.835$^{+0.016}_{-0.019}$ & 0.54 & 0.823$^{+0.018}_{-0.018}$ & 0.54 & 0.798$^{+0.032}_{-0.034}$ & 0.45\\
		\hline
	\end{tabular}
	\renewcommand{\arraystretch}{1.0}
	\label{tab:best_fit_confidence}
\end{table*}

\begin{table}
	\centering
	\caption{Similar to \tab{tab:criteria_confidence}, but for credible regions. The description of ABC can be found in \sect{sect:ABC}.}
	\renewcommand{\arraystretch}{1.3}
	\begin{tabular}{c|cc|cc|cc}
		\hline\hline
		          & \multicolumn{2}{c|}{$\bx^{\abd 5}$} & \multicolumn{2}{c|}{$\bx^{\pct 5}$} & \multicolumn{2}{c}{$\bx^{\cut 5}$}\\[-0.5ex]
		\hline
		          & $\Delta\Sigma_8$ & FoM & $\Delta\Sigma_8$ & FoM & $\Delta\Sigma_8$ & FoM\\[-0.3ex]
		\hline
		$L_\cg$   & 0.033            & 43  & 0.038            & 31  & 0.066            & 15 \\[-0.5ex]
		$L_\svg$  & 0.031            & 53  & 0.033            & 41  & 0.056            & 20 \\[-0.5ex]
		$L_\vg$   & 0.031            & 53  & 0.032            & 40  & 0.055            & 18 \\[-0.5ex]
		$L_\cc$   & 0.033            & 40  & 0.040            & 30  & 0.071            & 14 \\[-0.5ex]
		$L_\vc$   & 0.033            & 47  & 0.035            & 36  & 0.060            & 16 \\[-0.5ex]
		$L_\true$ & 0.034            & 49  & 0.036            & 36  & 0.061            & 17 \\[-0.5ex]
		ABC       & 0.056            & 31  & 0.044            & 33  & 0.068            & 16 \\
		\hline
	\end{tabular}
	\renewcommand{\arraystretch}{1.0}
	\label{tab:criteria_credible}
\end{table}

\begin{table*}
	\centering
	\caption{Similar to \tab{tab:best_fit_confidence}, but for Bayesian inference (credible region). The description of ABC can be found in \sect{sect:ABC}. We note that the best fits for $\Sigma_8$ are indicative since we do not use the real observational data in this study.}
	\renewcommand{\arraystretch}{1.3}
	\begin{tabular}{c|cc|cc|cc}
		\hline\hline
		          & \multicolumn{2}{c|}{$\bx^{\abd 5}$} & \multicolumn{2}{c|}{$\bx^{\pct 5}$} & \multicolumn{2}{c}{$\bx^{\cut 5}$}\\[-0.5ex]
		\hline
		          & $\Sigma_8$ $^{+1\sigma}_{-1\sigma}$ & $\alpha$ & $\Sigma_8$ $^{+1\sigma}_{-1\sigma}$ & $\alpha$ & $\Sigma_8$ $^{+1\sigma}_{-1\sigma}$ & $\alpha$\\[0.1ex]
		\hline
		$L_\cg$   & 0.831$^{+0.017}_{-0.016}$ & 0.54 & 0.822$^{+0.018}_{-0.020}$ & 0.54 & 0.800$^{+0.030}_{-0.036}$ & 0.45\\[0.2ex]
		$L_\svg$  & 0.831$^{+0.016}_{-0.015}$ & 0.52 & 0.820$^{+0.016}_{-0.017}$ & 0.51 & 0.800$^{+0.032}_{-0.024}$ & 0.40\\[0.2ex]
		$L_\vg$   & 0.829$^{+0.015}_{-0.015}$ & 0.52 & 0.819$^{+0.015}_{-0.017}$ & 0.52 & 0.800$^{+0.025}_{-0.029}$ & 0.42\\[0.2ex]
		$L_\cc$   & 0.830$^{+0.017}_{-0.017}$ & 0.54 & 0.825$^{+0.018}_{-0.022}$ & 0.54 & 0.807$^{+0.030}_{-0.041}$ & 0.46\\[0.2ex]
		$L_\vc$   & 0.829$^{+0.016}_{-0.016}$ & 0.52 & 0.823$^{+0.016}_{-0.019}$ & 0.53 & 0.798$^{+0.030}_{-0.030}$ & 0.44\\[0.2ex]
		$L_\true$ & 0.828$^{+0.019}_{-0.015}$ & 0.53 & 0.823$^{+0.015}_{-0.021}$ & 0.53 & 0.800$^{+0.030}_{-0.032}$ & 0.44\\[0.2ex]
		ABC       & 0.819$^{+0.030}_{-0.025}$ & 0.50 & 0.817$^{+0.022}_{-0.022}$ & 0.51 & 0.799$^{+0.034}_{-0.034}$ & 0.42\\
		\hline
	\end{tabular}
	\renewcommand{\arraystretch}{1.0}
	\label{tab:best_fit_credible}
\end{table*}

\subsection{$p$-value analysis}
\label{sect:nonParam_pValue}

\begin{figure}[tb]
	\centering
	\includegraphics[width=8.5cm]{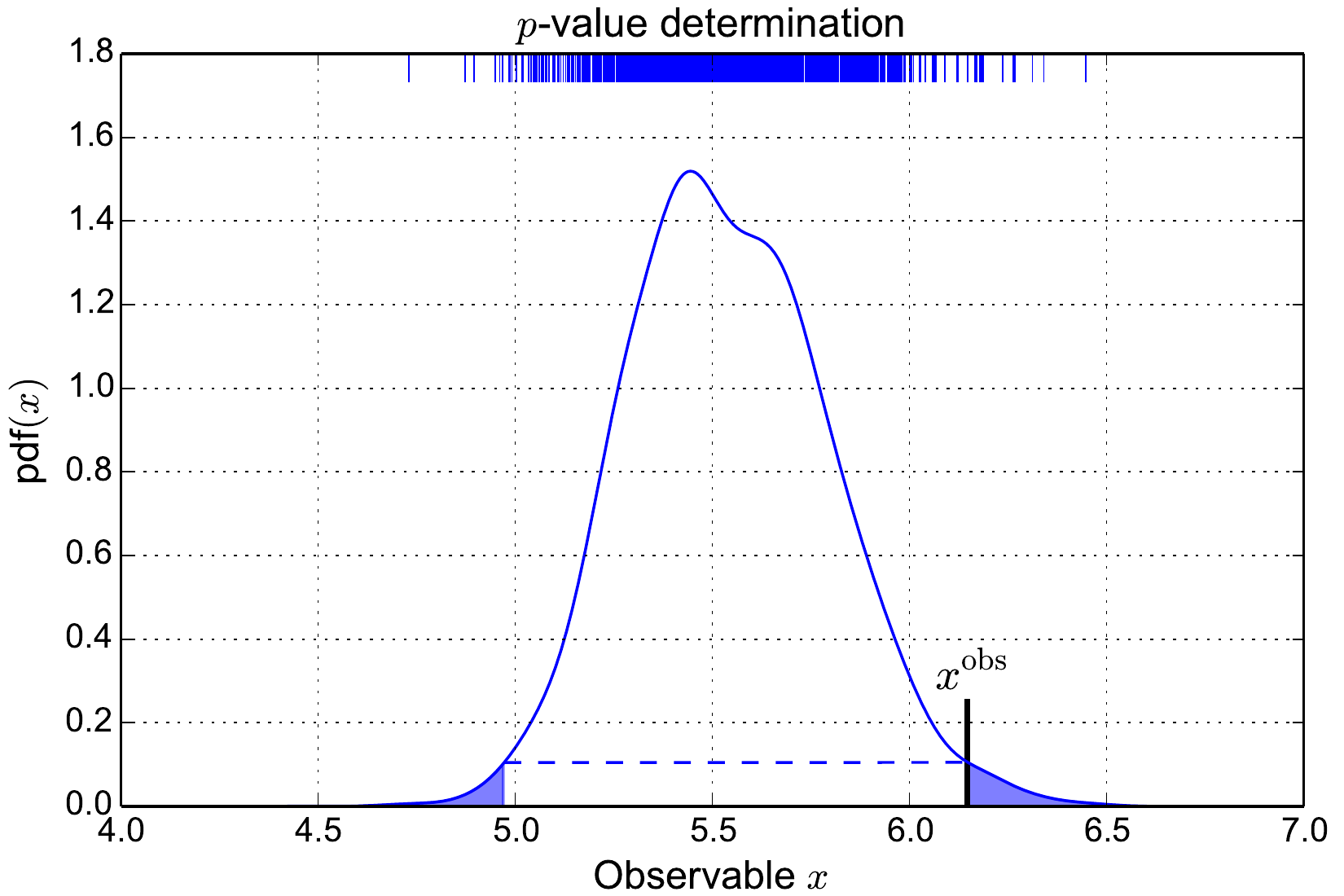}
	\caption{Example for the $p$-value determination. The $x$-axis indicates a one-dimensional observable, and the $y$-axis is the PDF. The PDF is obtained from a kernel density estimation using the $N = 1000$ realizations. Their values are shown as bars in the rug plot at the top of the panel. The shaded area is the corresponding $p$-value for given observational data $x^\obs$.}
	\label{fig:p_value_determination}
\end{figure}

\begin{figure}[tb]
	\centering
	\includegraphics[width=8.5cm]{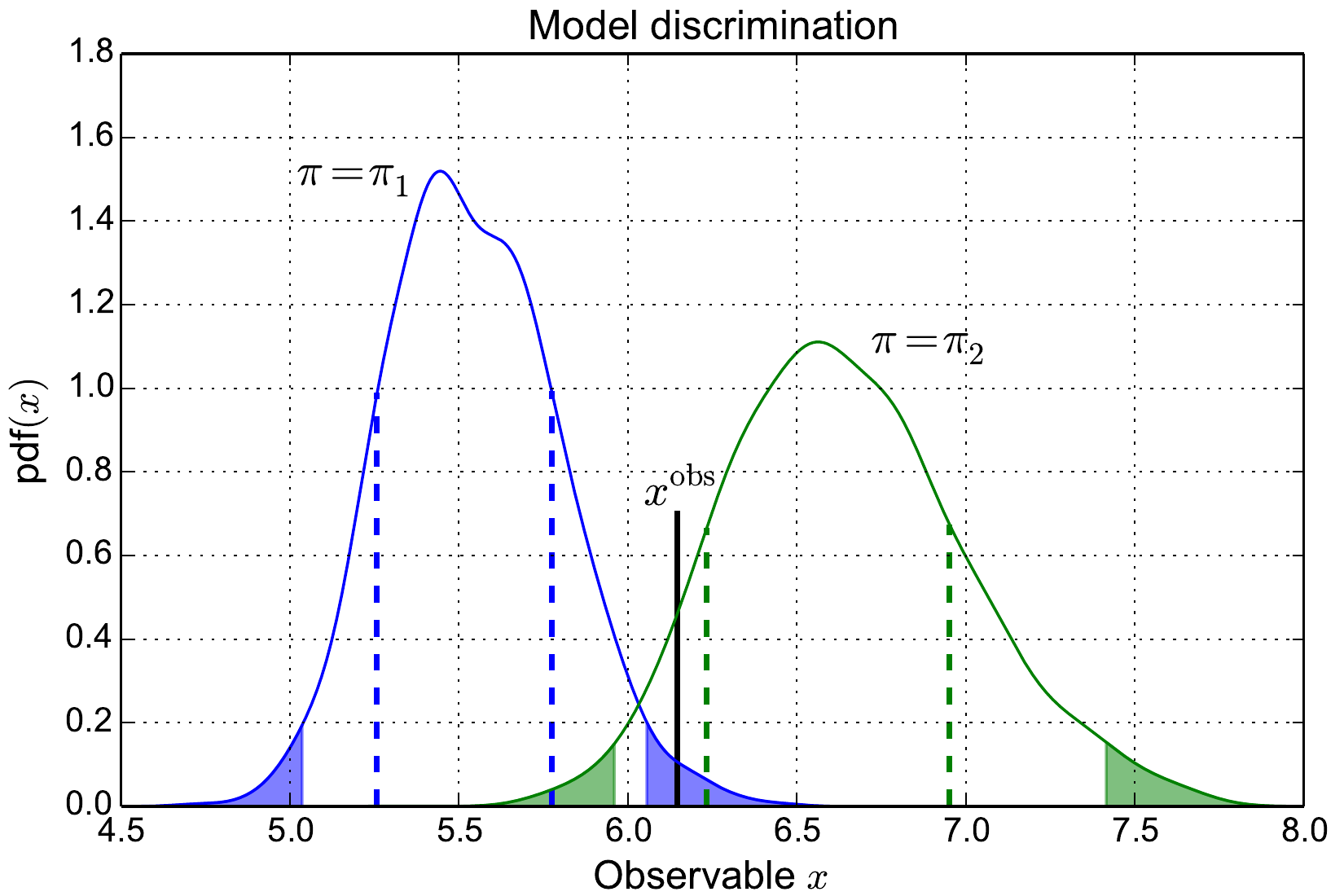}
	\caption{PDF from two different models (denoted by $\pi_1$ and $\pi_2$) and the observation $x^\obs$. The dashed lines show 1-$\sigma$ intervals for both models, while the shaded areas are intervals beyond the 2-$\sigma$ level. In this figure, model $\pi_1$ is excluded at more than 2-$\sigma$, whereas the significance of the model $\pi_2$ is between 1 and 2-$\sigma$.}
	\label{fig:model_discrimination}
\end{figure}
 
Another non-analytic technique is the $p$-value analysis. This frequentist approach provides the significance level by directly determining the $p$-value associated with a observation $\bx^\obs$. The $p$-value is defined as 
\begin{align}\label{for:p_value}
        p \equiv 1 - \int \rmd^d\bx\ \hat{P}(\bx|\bpi)\times \Theta\left(\hat{P}(\bx|\bpi)- \hat{P}(\bx^\obs|\bpi)\right),
\end{align}
where $\Theta$ denotes the Heaviside step function. The integral extends over the region where $\bx$ is more probable than $\bx^\obs$ for a given $\bpi$, as shown by \fig{fig:p_value_determination}. Thus, the interpretation of \for{for:p_value} is that if we generated $N$ universes, then at least $(1-p)N$ of them should have an observational result ``better'' than $\bx^\obs$. In this context, "better" refers to a more probable realization. The significance level is determined by the chi-squared distribution with $d=2$ degree of freedom, for two free parameters, $\Omegam$ and $\sigeig$. As \fig{fig:model_discrimination} shows, this provides a straightforward way to distinguish different cosmological models.

As in \sect{sect:copula_constraints}, we used KDE to estimate the multivariate PDF and numerically integrated \for{for:p_value} to obtain the $p$-value. Monte Carlo integration is used for evaluating the five-dimensional integrals.

\subsection{Parameter constraints}
\label{sect:nonParam_constraints}

\begin{figure*}[tb]
	\centering
	\includegraphics[width=8.5cm]{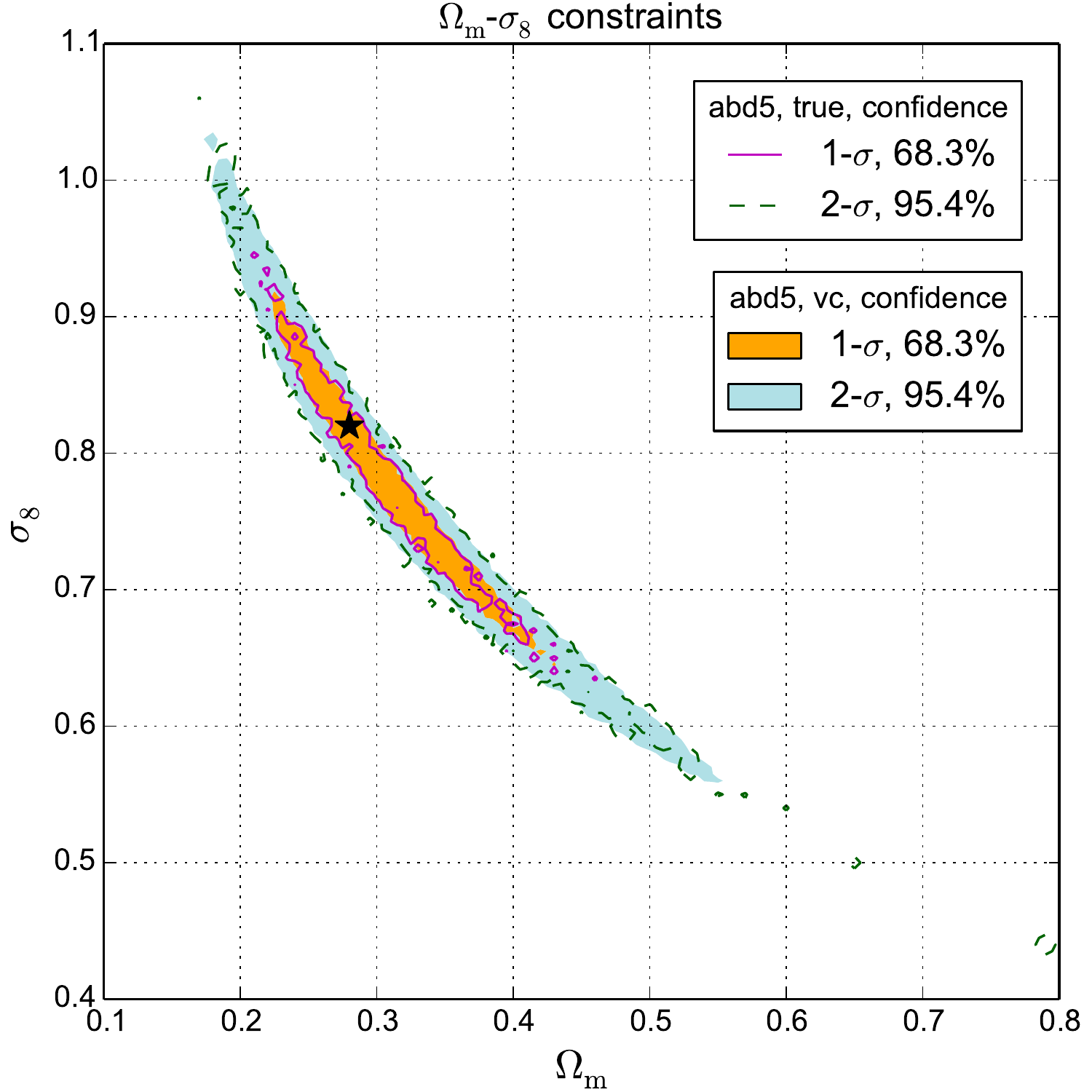}
	\includegraphics[width=8.5cm]{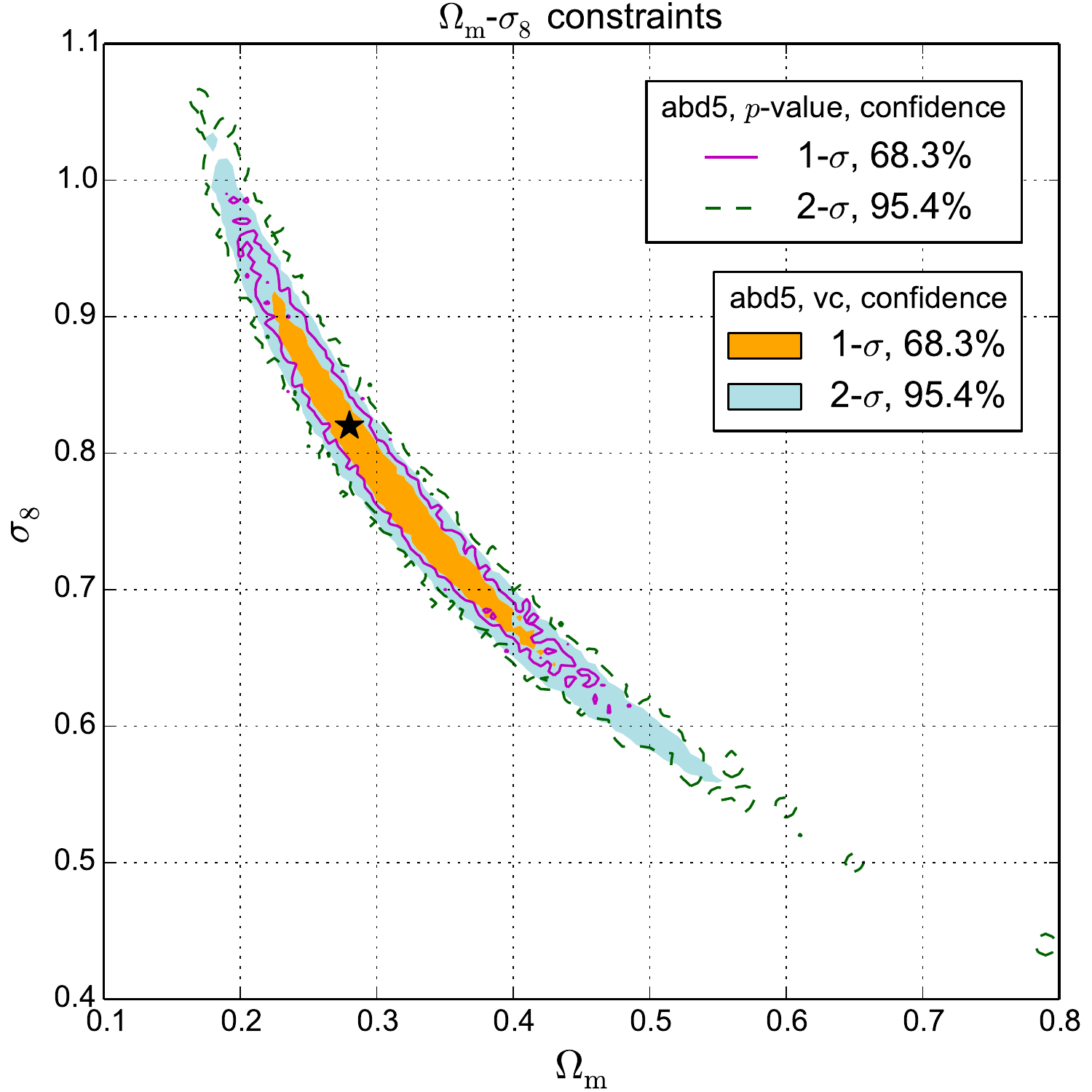}

	\caption{\textit{Left panel}: confidence regions derived from $L_\vc$
(colored areas) and $L_\true$ (solid and dashed lines) with $\bx^{\abd 5}$.
\textit{Right panel}: confidence regions derived from
$L_\vc$ (colored areas) and $p$-value analysis (solid and dashed lines). The
contours from $L_\true$ and $p$-value analysis are noisy due to a relatively
low $N$ for probability estimation. We notice that $L_\vc$ and $L_\true$ yield
very similar results.}

	\label{fig:contour_real_abd5}
\end{figure*}

\figFull{fig:contour_real_abd5} shows the confidence contours from $L_\true$
and $p$-value analysis with observables $\bx^{\abd 5}$. We notice that these
constraints are very noisy. This is due to a relatively low number of
realizations to estimate the probability and prevents us from making definite
conclusions. Nevertheless, the result from the lefthand panel reveals good
agreement between constraints from two likelihoods. This suggests that we may
substitute $L_\true$ with the CDC-copula likelihood to bypass the drawback of
noisy estimation from $L_\true$. In the righthand panel, the
result from $p$-value analysis seems to be larger. We
reduced the noise for the $p$-value analysis by combining $\bx^{\abd 5}$ into a
two-component vector. In this case, the $p$-value is evaluated using the grid
integration. This data compression technique does not significantly enlarge 
but visibly smooths the contours.

In the $L_\true$ case, the probability information that we need is local
since the likelihood $P$ is only evaluated at $\bx^\obs$. For $p$-value
analysis, one needs to determine the region where $P(\bx) < P(\bx^\obs)$ and
integrate over it, so a more global knowledge of $P(\bx)$ is needed in this
case. We recall that KDE smooths, thus the estimation is always
biased \citep{Zambom_Dias_2012}. Other estimations, for example using the
Voronoi-Delaunay tessellation \citep[see, e.g.,][]{Schaap_2007,
Guio_Achilleos_2009}, could be an alternative to the KDE technique. As a
result, observable choice, data compression, and density estimation need to be
considered jointly for all non-analytic approaches.

Recent results from CFHTLenS
\citepalias{Liu_etal_2015} and Stripe-82 
\citepalias{Liu_etal_2015a} resulted in $\Delta\Sigma_8\sim
0.1$, about 2--3 times larger than this study. However, we would like to
highlight that redshift errors are not taken into account here and that the
simulated galaxy density used in this work is much higher. Also, we choose $z_\rms = 1,$
which is higher than the median redshift of both surveys ($\sim 0.75$). All
these factors contribute to our smaller error bars.

\section{Approximate Bayesian computation}
\label{sect:ABC}

\subsection{PMC ABC algorithm}
\label{sect:ABC_algorithm}

In the previous section, we presented parameter constraints derived from
directly evaluating the underlying PDF. Now, we want to move a step
further and bypass the likelihood estimation altogether. 

Based on
an accept-reject rule, approximate Bayesian computation (ABC) is an
algorithm that provides an approximate posterior distribution of a complex stochastic
process when evaluating the likelihood is expensive or unreachable. There
are only two requirements: (1) a stochastic model for the observed data that
samples the likelihood function of the observable and (2)
a measure, called summary statistic, to perform model comparison. We
present below a brief description of ABC.
Readers can find detailed reviews of ABC in \citet{Marin_etal_2011} and
Sect.~1 of \citet{Cameron_Pettitt_2012}.

The idea behind ABC can be most easily illustrated in the case of discrete
data as follows. Instead of explicitly calculating the likelihood, one first
generates a set of parameters $\{\bpi_i\}$ as samples under the prior
$\mathcal{P}(\bpi)$, and then for each $\bpi_i$ simulates a model prediction $\bX$
sampled under the likelihood function $P(\cdot|\bpi_i).$ (Here we put $\bX$
in the upper case to emphasize that $\bX$ is a random variable.) Keeping only
those $\bpi_i$ for which $\bX=\bx^\obs$, the distribution of the accepted
samples $\mathcal{P}_\mathrm{ABC}(\bpi)$ equals the posterior distribution of the
parameter $\mathcal{P}(\bpi|\bx^\obs)$ given the observed data, since
\begin{align}
  	\mathcal{P}_\mathrm{ABC}(\bpi) &= \sum_{\bX} P(\bX|\bpi) \mathcal{P}(\bpi) \delta_{\bX, \bx^\obs} \notag\\
  	&= P(\bx^\obs|\bpi) \mathcal{P}(\bpi) \notag\\
  	&= \mathcal{P}(\bpi|\bx^\obs),
  	\label{eq:ABC_post_discrete}
\end{align}
where $\delta_{\bX, \bx^\obs}$ is Kronecker's delta. Therefore, $\{\bpi_i\}$ is an independent and identically distributed sample from the posterior.
It is sufficient to perform a \emph{one-sample test}: using a single realization $\bX$ for each parameter
$\pi$ to obtain a sample under the posterior.

ABC can also be adapted to continuous data and parameters, where obtaining a strict equality
$\bX=\bx^\obs$ is pratically impossible. As a result, sampled points are
accepted with a \emph{tolerance level} $\epsilon$, say
$|\bX-\bx^\obs|\leq\epsilon$.
What is retained after repeating this process
is an ensemble of parameters $\bpi$ that are compatible with the data 
and that follow
a probability distribution, which is a modified version of \for{eq:ABC_post_discrete},
\begin{align}\label{for:ABC_posterior}
	\mathcal{P}_\epsilon(\bpi|\bx^\obs) = A_\epsilon(\bpi) \mathcal{P}(\bpi),
\end{align}
where $A_\epsilon(\bpi)$ is the probability that a proposed parameter
$\bpi$ passes the one-sample test within the error $\epsilon$:
\begin{align}\label{for:ABC_accept}
	A_\epsilon(\bpi) \equiv \int \rmd \bX\ P(\bX | \bpi)\mathbbm{1}_{|\bX-\bx^\obs|\leq\epsilon}(\bX).
\end{align}
The Kronecker delta from \for{eq:ABC_post_discrete} has now
been replaced with the indicator function $\mathbbm{1}$ of the set of 
points $\bX$ that satisfy the tolerance criterion. The basic assumption of ABC is 
that the probability distribution (\ref{for:ABC_posterior}) is a good approximation
of the underlying posterior, such that
\begin{equation}
	\mathcal{P}_\epsilon(\bpi|\bx^\obs) \approx \mathcal{P}(\bpi|\bx^\obs).
	\label{for:ABC_approximation}
\end{equation}
Therefore, the error can be seperated into two parts:
one from the approximation above and the other
from the estimation of the desired integral, $\mathcal{P}_\epsilon$.
For the latter, gathering one-sample tests of $A_\epsilon$ makes 
a Monte Carlo estimation of $\mathcal{P}_\epsilon$, which is unbiased.
This ensures the use of the one-sample test.

A further addition to the ABC algorithm is a reduction in the complexity of the
full model and data. This is necessary in cases of very large dimensions,
for example, when the model produces entire maps or large catalogs. The
reduction of data complexity is done with the so-called \emph{summary
statistic} $s$. For instance, in our peak-count framework, a complete data set
$\bx$ is a peak catalog with positions and S/N values, and the summary
statistic $s$ is chosen here to be $s(\bx) = \bx^{\abd 5}$, $\bx^{\pct 5}$, or
$\bx^{\cut 5}$, respectively, for the three cases of observables introduced
in Sect.~\ref{sect:methodology_design}. As a remark, if this 
summary statistic is indeed \emph{sufficient}, then \for{for:ABC_approximation} 
will no longer be an approximation. The true posterior can be recovered when
$\epsilon \rightarrow 0$.

For a general comparison of model and data, one chooses a metric $D$ adapted to
the summary statistic $s$, and the schematic expression
$|\bX-\bx^\obs|\leq \epsilon$ used above is generalized to $D(s(\bX),
s(\bx^\obs))\leq\epsilon$. We highlight that the summary statistic can have a low
dimension and a very simple form. In practice, it is motivated by
computational efficiency, and it seems that a simple summary can still
produce reliable constraints
\citep{Weyant_etal_2013}.

The integral of \for{for:ABC_posterior} over $\bpi$ is smaller than unity, and
the deficit only represents the probability that a parameter is rejected by
ABC. This is not problematic since density estimation will automatically
normalize the total integral over the posterior. However, a more subtle issue
is the choice of the tolerance level $\epsilon$. If $\epsilon$ is too high,
$A(\bpi)$ is close to $1$, and \for{for:ABC_approximation} becomes a bad
estimate. If $\epsilon$ is too low, $A(\bpi)$ is close to $0$, and sampling
becomes extremely difficult and inefficient. How, then, should one choose
$\epsilon$? This can be done by applying the iterative importance sampling
approach of population Monte Carlo \citep[PMC; for applications to cosmology,
see][]{Wraith_etal_2009} and combine it with ABC \citep{DelMoral_etal_2006,
Sisson_etal_2007}. This joint approach is sometimes called SMC ABC,
where SMC stands for sequential Monte Carlo; we refer to it as PMC ABC.
The idea of PMC ABC is to iteratively reduce the tolerance $\epsilon$ until a
stopping criterion is reached.

\begin{algorithm}[tb]
	\caption{Population Monte Carlo approximate Bayesian computation}
	\label{algo:ABC}
	\begin{algorithmic}
		\REQUIRE
			\STATE number of particles $p$
			\STATE prior $\rho(\cdot)$
			\STATE summary statistic $s(\cdot)$
			\STATE distance $D(\cdot, \cdot)$
			\STATE shutoff parameter $r_\mathrm{stop}$
		\bigskip
		\STATE set $t = 0$
		\FOR{$i = 1$ to $p$}
			\STATE generate $\bpi_i\upp{0}$ from $\rho(\cdot)$ and $\bx$ from $\pdf\left(\cdot|\bpi_i\upp{0}\right)$
			\STATE set $\delta_i\upp{0} = D \left(s(\bx), s(\bx^\obs)\right)$ and $w_i\upp{0} = 1/p$
		\ENDFOR
		\STATE set $\epsilon\upp{1} = \median\left(\delta_i\upp{0}\right)$ and $\bC\upp{0} = \cov\left(\bpi_i\upp{0}, w_i\upp{0}\right)$ 
		\bigskip
		\WHILE{success rate $\geq r_\mathrm{stop}$}
			\STATE $t \leftarrow t + 1$
			\FOR{$i = 1$ to $p$}
				\REPEAT 
					\STATE generate $j$ from {\footnotesize $\{1,\ldots, p\}$} with weights {\footnotesize $\{w_1\upp{t-1}, \ldots, w_p\upp{t-1}\}$}
					\STATE generate $\bpi_i\upp{t}$ from {\footnotesize $\mathcal{N}\left(\bpi\upp{t-1}_j, \bC\upp{t-1}\right)$} and $\bx$ from $\pdf\left(\cdot|\bpi_i\upp{t}\right)$
					\STATE set $\delta_i\upp{t} = D \left(s(\bx), s(\bx^\obs)\right)$
				\UNTIL{$\delta_i\upp{t} \leq \epsilon\upp{t}$}
				\STATE set $w_i\upp{t} \propto \displaystyle\frac{\rho\left(\bpi_i^{(t)}\right)}{\sum_{j=1}^P w_j\upp{t-1} K\left(\bpi\upp{t}_i-\bpi\upp{t-1}_j, \bC\upp{t-1}\right)}$
			\ENDFOR
			\STATE set $\epsilon\upp{t+1} = \median\left(\delta_i\upp{t}\right)$ and $\bC\upp{t} = \cov\left(\bpi_i\upp{t}, w_i\upp{t}\right)$ 
		\ENDWHILE
	\end{algorithmic}
\end{algorithm}

Algorithm \ref{algo:ABC} details the steps for PMC ABC. 
We let $\pdf(\bx|\bpi)$ be a probabilistic model for $\bx$
given $\bpi$. PMC ABC requires a prior $\rho(\bpi)$, a summary statistic
$s(\bx)$ that retains only partial information about $\bx$, a distance
function $D$ based on the dimension of $s(\bx)$, and a shutoff parameter
$r_\mathrm{stop}$. We denote $\mathcal{N}(\bpi, \bC)$ as a
multivariate normal with mean $\bpi$ and covariance matrix $\bC$, $K(\bpi,
\bC) \propto \exp\left( -\bpi^{\rm T}\bC\inv\bpi /2\right)$ a Gaussian kernel,
$\cov(\bpi_i, w_i)$ the weighted covariance for the set $\{\bpi_1, \ldots,
\bpi_p\}$ with weights $\{w_1, \ldots, w_p\}$, and $p$ the number of
\emph{particles}, i.e.,~the number of sample points in the parameter space.

In the initial step, PMC ABC accepts all particles drawn from the prior and
defines an acceptance tolerance before starting the first iteration. The
tolerance is given by the median of the distances of the summary statistic
between the observation and the stochastic model generated from each particle.
Then, each iteration is carried out by an importance-sampling step based on
weights determined by the previous iteration. To find a new particle, a
previous point $\bpi_j\upp{t-1}$ is selected according to its weight. A
candidate particle is drawn from a proposal distribution, which is a normal law
centered on $\bpi_j\upp{t-1}$ with a covariance equal to the covariance of all
particles from the previous iteration. With a model generated using the
candidate particle,  we accept the new particle if the distance between the
model and the observation is shorter than the tolerance, and reject it
otherwise. After accepting $p$ particles, the success rate, defined as the
ratio of accepted particles to total tries, is updated. The iterations continue
until the success rate decreases below the shutoff value. Instead of defining a
minimal tolerance \citep{Beaumont_etal_2002, Fearnhead_Prangle_2010,
Weyant_etal_2013}, we use a simplified stopping criterion that is based on the
selection efficiency of the algorithm. Since \cite{McKinley_etal_2009} prove
that the stopping criterion has very little impact on the estimated posterior,
the choice of the tolerance level is instead a question of computational power.

\subsection{Settings for ABC}
\label{sect:ABC_settings}

\cite{McKinley_etal_2009} studied the impact of the various choices necessary for ABC,
by comparing Markov chain Monte Carlo (MCMC) ABC and PMC ABC.
The authors concluded that
(1) increasing the number of simulations beyond one for a
given parameter does not seem to improve the posterior estimation 
\citep[similar conclusion found by][]{Bornn_etal_2014},
(2) the specific choice of the tolerance level does not seem to be important,
(3) the choice of the summary statistic and the distance is crucial, and
(4) PMC ABC performs better than MCMC ABC.
Therefore, exploring a sufficient summary statistic to represent the whole data
set becomes an essential concern for the ABC technique.

To solve the optimal data compression problem, \cite{Blum_etal_2013} provide a series of methods
in their review for selecting the ideal summaries, as well as
methods of reducing the data dimension. Leaving a detailed study of optimal choice
for the future work, we adopted a straightforward summary statistic in this work,
defined as $s(\bx) = \bx^{\mathrm{type}5}$, and the distance as
\begin{align}
  D(\bx^{\mathrm{type}5}, \vect{y}^{\mathrm{type}5}) = \sqrt{\sum_{i=1}^5
  \frac{\left( x^{\mathrm{type}5}_i-y^{\mathrm{type}5}_i \right)^2}{C_{ii}}}.
\end{align}
This is simply a weighted Euclidean distance, where the weight
$C_{ii}\inv$ is needed to level out the values of the different S/N.

The prior $\rho$ is chosen to be flat. We set $p = 250$ and $r_\mathrm{stop} =
0.02$. In this condition, we can easily compare the computational cost with the
analyses presented in previous sections. If $\tau$ is the time cost for one
model realization, the total time consumption is $7821\times 1000\times \tau$
for our likelihood-based analyses in Sects. \ref{sect:CDC}, \ref{sect:copula},
and \ref{sect:nonParam}, and $250 \times 1 \times (\sum_t r_t\inv) \times \tau$
for ABC where $r_t$ is the acceptance rate of the $t$-th iteration. For
$\bx^{\abd 5}$, $\sum_t r_t\inv \approx 102$ ($0\leq t\leq 9$), so the
computation time for ABC is drastically reduced by a factor of $\sim 300$
compared to the likelihood analyses. ABC is faster by a similar
factor compared to Monte-Carlo sampling, since typically the number of required
sample points is ${\cal O}(10^4)$, the same order of magnitude as our number of
grid points.

\subsection{Results from ABC}
\label{sect:ABC_results}

\begin{figure*}[tb]
	\centering
	\includegraphics[width=17cm]{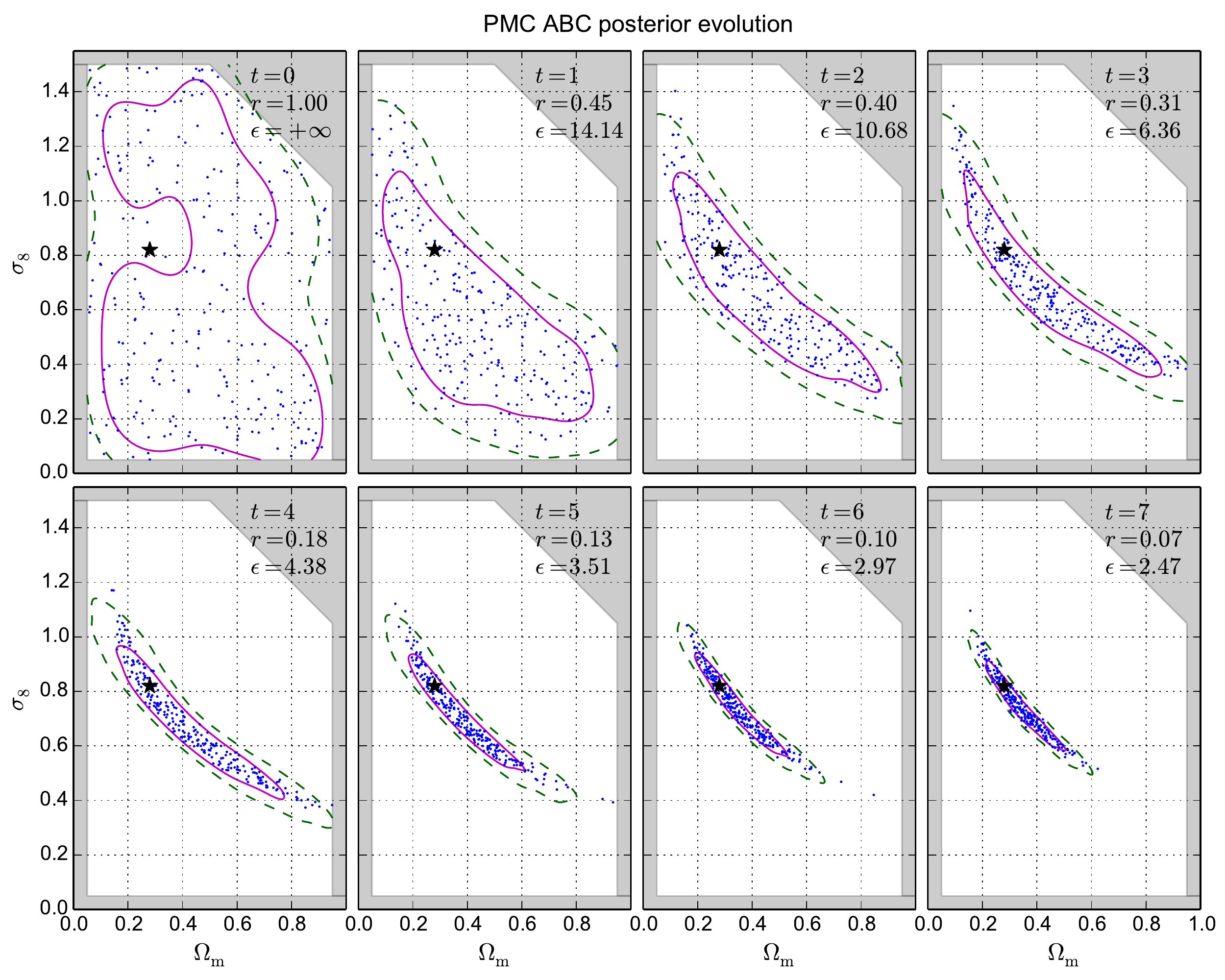}
	\caption{Evolution of particles and the posterior from the PMC ABC algorithm. We show results from the first 8 iterations ($0\leq t \leq 7$). Particles are given by blue dots. Solid lines are 1-$\sigma$ contours, and dashed lines are 2-$\sigma$ contours. White areas represent the prior. The corresponding accept rate $r$ and tolerance level $\epsilon$ are also given. We set $\epsilon\upp{0} = +\infty$.}
	\label{fig:ABC_evolution_abd5}
\end{figure*}

\figFull{fig:ABC_evolution_abd5} shows the iterative evolution of the PMC ABC
particles. We drew the position of all 250 particles and credible regions for
the first eight iterations. The summary statistic is $\bx^{\abd 5}$. The credible
regions were drawn from the posterior estimated on a grid using KDE with the ABC
particles as sample points $\bx^{(k)}$ in \for{for:multivariate_KDE}.
We ignored the particle weights for this density estimate.
We find that the contours stablize for $t \geq 8$, which
correponds to an acceptance rate of $r=0.05$. At these low accpetence rates,
corresponding to a small tolerance, the probability of satisfying the tolerance
criterion $D(s(\bX), s(\bx^\obs))\leq\epsilon$ is low even though
$\bX$ is sampled from parameters in the high-probability region, and
accepting a proposed particle depends mainly on random fluctuations due to the
stochasticity of the model.

\begin{figure}[tb]
	\centering
	\includegraphics[width=8.5cm]{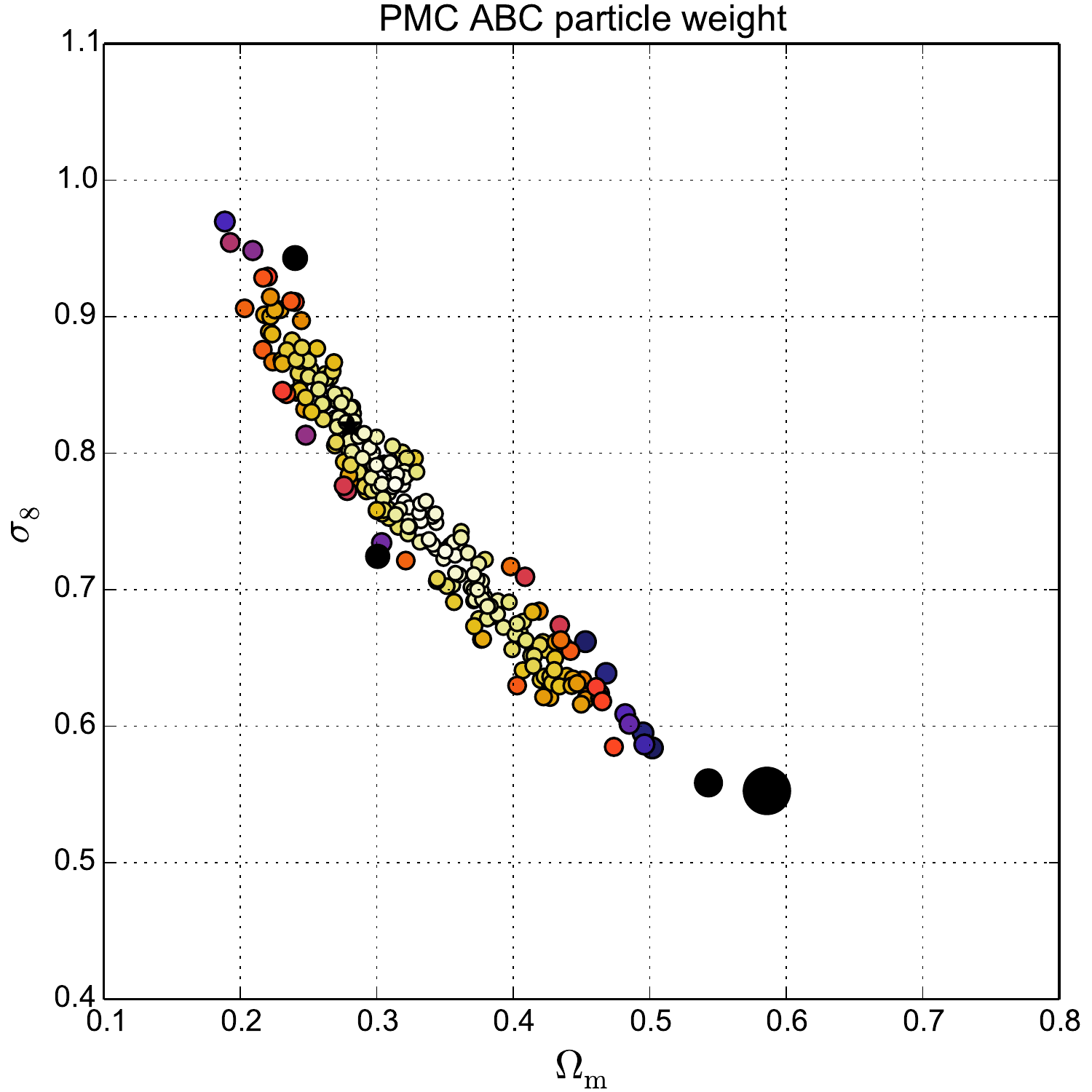}
	\caption{Weights of particles from $t=8$ with $s(\bx) = \bx^{\abd 5}$. The weight is represented by the size and the color at the same time.}
	\label{fig:ABC_weight_abd5_p250_t8}
\end{figure}

\begin{figure}[tb]
	\centering
	\includegraphics[width=8.5cm]{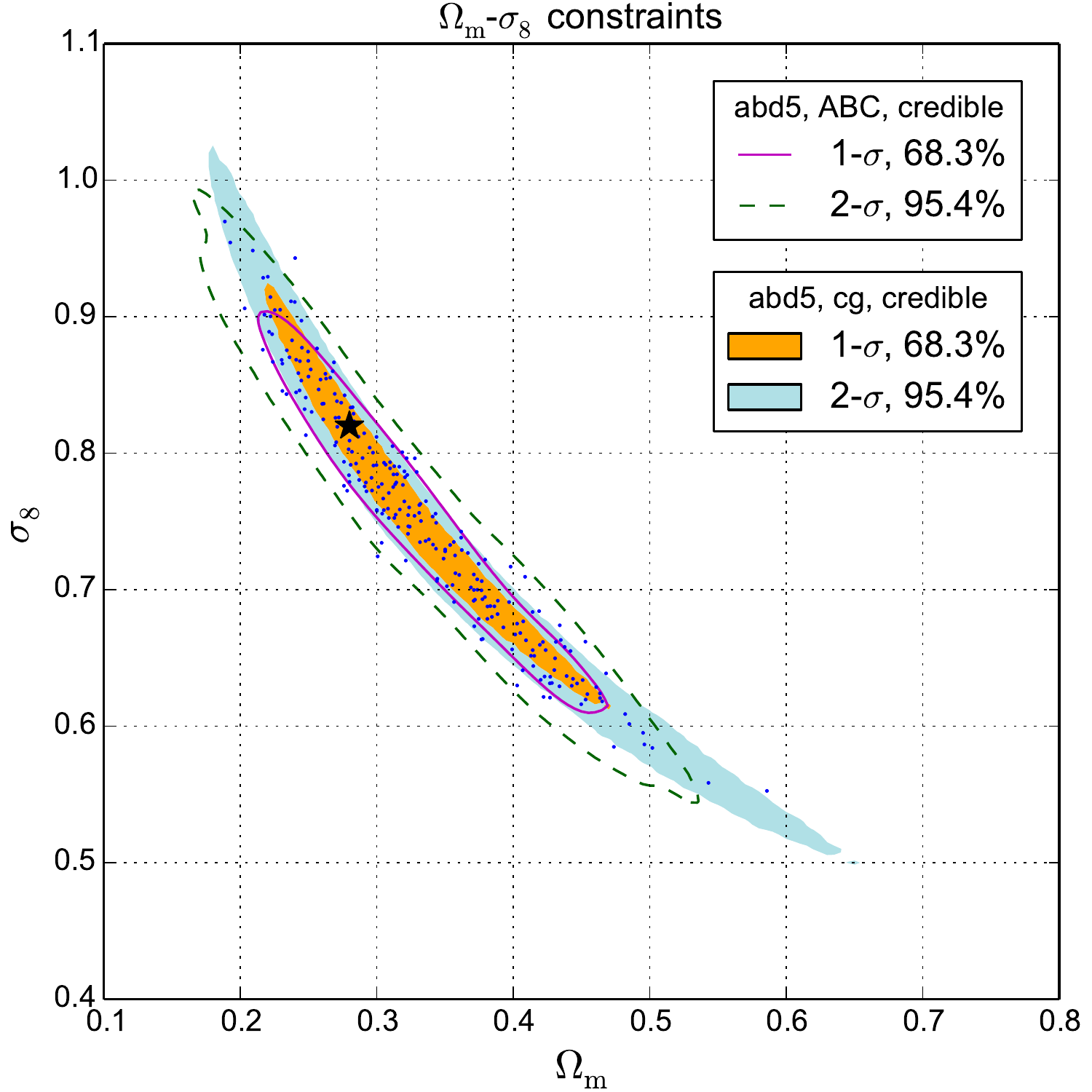}
	\caption{Comparison between credible regions derived from $L_\cg$ (colored areas) and ABC (solid and dashed lines).}
	\label{fig:contour_abd5_cg_ABC_credible}
\end{figure}

In \fig{fig:ABC_weight_abd5_p250_t8}, we show the weights of particles sampled
at the final iteration $t=8$. The weight is visualized by both color and
size of the circle. The figure shows that points
farther away from the maximum have larger weights as constructed by
Algorithm \ref{algo:ABC}. Since those points are more isolated, their weights compensate
for the low density of points, avoiding undersampling the
tails of the posterior and overestimating the
constraining power.

In \fig{fig:contour_abd5_cg_ABC_credible} we show the comparison of credible
regions between $L_\cg$ and PMC ABC with $s(\bx) = \bx^{\abd 5}$. The FoM and
the best-fit ABC results are presented in Tables \ref{tab:criteria_credible}
and \ref{tab:best_fit_credible}, respectively. The figure shows good
agreement between the two cases, and thus validates the performance of PMC ABC.
The broader contours from ABC might be caused by a bias of KDE. 
The same reason might be responsible for the slight shift of the contours in
the tails of the distribution, which do not follow the particles exactly, which are best
visible in the two lefthand panels in the lower row of
\fig{fig:ABC_evolution_abd5}.

\section{Summary and discussion}
\label{sect:summary}

Our model for weak-lensing peak counts, which provides a direct estimation of the underlying PDF of observables, leads to a wide range of possibilities for constraining parameters. To summarize this work, we
\begin{itemize}
	\item compared different data vector choices,
	\item studied the dependence of the likelihood on cosmology,
	\item explored the full PDF information of observables,
	\item proposed different constraint strategies, and
	\item examined them with two criteria.
\end{itemize}

In this paper, we performed three different series of analyses--the Gaussian
likelihood, the copula likelihood, and non-analytic analyses--by using three
different data vectors: one based on the peak PDF and two on the
CDF. We defined two quantitative criteria: $\Delta\Sigma_8$, which
represents the error bar on the parameter $\Sigma_8 =
\sigeig(\Omegam/0.27)^\alpha$ and is a measure of the width of the
$\Omegam$-$\sigeig$ degeneracy direction; and FoM, which is the area of the
$\Omegam$-$\sigeig$ contour. Both Bayesian and frequentist approaches were followed. Although the interpretations are different, the results are
very similar.

We studied the cosmology-dependent-covariance (CDC) effect by estimating the true covariance for each parameter set. We found that the CDC effect can increase the constraining power up to 22\%. The main contribution comes from the additional variation of the $\chi^2$ term, and the contribution from the determinant term is negligible. These observations confirm a previous study by \cite{Eifler_etal_2009}.

We also performed a copula analysis, which makes weaker assumptions than
Gaussianity. In this case, the marginalized PDF is Gaussianized by the copula
transform. The result shows that the difference with the Gaussian likelihood is
small. This is dominated by the CDC effect if a varying covariance
is taken into account.

Discarding the Gaussian hypothesis on the PDF of observables, we provided two
straightforward ways of using the full PDF information. The first one is the true
likelihood. The direct evaluation of the likelihood is
noisy owing to the high statistical fluctuations from the finite
number of sample points. However, we find that the varying-covariance copula
likelihood, noted as $L_\vc$ above, seems to be a good approximation to the
truth. The second method is to determine the $p$-value for a given
parameter set directly, and this approach gives us more conservative
constraints. We outline that both methods are covariance-free, avoiding
non-linear effects caused by the covariance inversion.

At the end we showed how approximate Bayesian computation (ABC) derives
cosmological constraints using the accept-reject sampling. Combined with
importance sampling, this method requires less computational resources than
all the others. We proved that by reducing the computational time by a factor of
300, ABC is able to yield consistent constraints from weak-lensing
peak counts. Furthermore, \cite{Weyant_etal_2013} show in their study that
ABC is able to perform unbiased constraints using contaminated data, 
demonstrating the robustness of this algorithm.

A comparison between different data vectors is done in this study. Although we
find for all analyses that $\bx^{\abd 5}$ outperforms $\bx^{\pct 5}$ by 
20\%--40\% in terms of FoM, this is not necessarily true in
general when we use a different percentile choice. Actually, the performance of
$\bx^\pct$ depends on the correlation between its different components.
However, the $\bx^\pct$ family is not recommended in practice because of model
biases induced for very low peaks (S/N <~0). In addition, our study shows that
the $\bx^\cut$ family is largely outperformed by $\bx^\abd$. Thus, we conclude
that $\bx^\abd$ seems to be good candidates for peak-count analysis, while the
change in the contour tilt from $\bx^\cut$ could be interesting when combining with
other information.

The methodology that we show for parameter constraints can be applied to all
fast stochastic forward models. Flexible and efficient, this approach possesses
a great potential whenever the modeling of complex effects is
desired. Our study displays two different parameter-constraint philosophies.
On the one hand, parameteric estimation (Sects. \ref{sect:CDC} and
\ref{sect:copula}), under some specific hypotheses such as Gaussianity,
only requires some statistical quantities such as the covariances. 
However, the appropriateness of the likelihood should
be examined and validated to avoid biases. On the other hand, non-analytic
estimation (Sects. \ref{sect:nonParam} and \ref{sect:ABC}) is directly derived
from the PDF. The problem of inappropriateness vanishes, but instead the
uncertainty and bias of density estimation become drawbacks. Depending on
modeling pertinence, an aspect may be more advantageous than another. Although not
studied in this work, a hybrid approach using semi-analytic estimator could be
interesting. This solicits more detailed studies of trade-off between the
inappropriatenss of analytic estimators and the uncertainty of density
estimation.

\begin{acknowledgements}
	This work is supported by R\'egion d'\^Ile-de-France in the framework of a DIM-ACAV thesis fellowship. We also acknowledge the support from the French national program for cosmology and galaxies (PNCG). The authors wish to acknowledge the anonymous referee for reviewing the paper. Chieh-An Lin would like to thank Karim Benabed, Ewan Cameron, Yu-Yen Chang, Yen-Chi Chen, C\'ecile Chenot, Fran\c{c}ois Lanusse, Sandrine Pires, Fred Ngol\`e, Florent Sureau, and Michael Vespe for useful discussions and suggestions on diverse subjects.
\end{acknowledgements}

\bibliography{Bibliographie_Linc}

\end{document}